# The STARFLAG handbook on collective animal behaviour: Part I, empirical methods

ANDREA CAVAGNA[1,2], IRENE GIARDINA[1,2], ALBERTO ORLANDI[1], GIORGIO PARISI[1,3], ANDREA PROCACCINI[1,3], MASSIMILIANO VIALE[3], VLADIMIR ZDRAVKOVIC[1]

[1] *Centre for Statistical Mechanics and Complexity (SMC), CNR-INFM*
[2] *Istituto dei Sistemi Complessi (ISC), CNR*
[3] *Dipartimento di Fisica, Universita' di Roma 'La Sapienza'*

Correspondence to Andrea Cavagna:
SMC, CNR-INFM, Dipartimento di Fisica, Universita' di Roma 'La Sapienza', Piazzale Aldo Moro 2, 00185 Roma, Italy
andrea.cavagna@roma1.infn.it

Irene Giardina, Alberto Orlandi, and Vladimir Zdravkovic:
SMC, CNR-INFM, Dipartimento di Fisica, Universita' di Roma 'La Sapienza', Piazzale Aldo Moro 2, 00185 Roma, Italy

Giorgio Parisi, Andrea Procaccini and Massimiliano Viale:
Dipartimento di Fisica, Universita' di Roma 'La Sapienza', Piazzale Aldo Moro 2, 00185 Roma, Italy

Massimiliano Viale (current address):
Dipartimento di Fisica, Universita' di Roma 3, via della Vasca Navale 84, 00146 Roma, Italy

## Abstract

The most startling examples of collective animal behaviour are provided by very large and cohesive groups moving in three dimensions. Paradigmatic examples are bird flocks, fish schools and insect swarms. However, because of the sheer technical difficulty of obtaining 3D data, empirical studies conducted to date have only considered loose groups of a few tens of animals. Moreover, these studies were very seldom conducted in the field. Recently the STARFLAG project achieved the 3D reconstruction of individual starlings' positions in cohesive flocks consisting of thousands of birds under field conditions, thus opening the way to a new generation of quantitative studies of collective animal behaviour. Here, we review the main technical problems in 3D data collection of large animal groups and we outline some of the methodological solutions adopted by the STARFLAG project. In particular, we explain how to solve the stereoscopic correspondence - or matching - problem, which was the major bottleneck of all 3D studies in the past.

**Keywords**: collective animal behaviour; self-organization; bird flocking; fish schooling; swarms; stereo photography.

The first speculations about collective animal behaviour date back to the observations of Pliny about flocks of starlings (translated by Rackham 1933). His remarks were necessarily very qualitative, although very reasonable too. Most of the hypotheses formulated almost two thousands years later were based on equally qualitative observations (Selous 1931; Emlen 1952). At times, the study of collective animal behaviour has been complemented by analogies with instances of collective behaviour in other fields of science, prominently physics (Radokov 1973). However, in the absence of any quantitative empirical insight, attempts to address the fundamental issues of collective behaviour rapidly became a matter of pure speculation.

It was only in the mid 1960s that the first empirical studies of collective animal behaviour in three dimensions led to some quantitative results. Cullen and co-workers studied small groups of up to 10 fish in a tank using the shadow method for three-dimensional (3D) reconstruction (Cullen et al. 1965). The same technique was later employed to study groups of up to 30 fish (Pitcher & Partridge 1979; Partridge et al. 1980; Partridge 1980). For the first time it was possible to measure interesting quantities, like the average nearest neighbour distance among individuals, and their angular distribution. Several other studies of fish groups have been performed since (see, for example, Van Long et al. 1985), although always with very small groups.

Empirical 3D data for birds have been harder to come by, because working in the field imposes serious constraints on accuracy. The first studies (Sugg 1965; van Tets 1966) focused on two-dimensional estimates of flocks' densities using a single photograph technique. It was only in 1978 that Major and Dill, using stereometry, could reconstruct the 3D positions of individual birds in small flocks of dunlins and starlings (on average 55 individuals per flock) (Major & Dill 1978). Subsequently, Pomeroy and Heppner (1992) managed to reconstruct individual birds' trajectories in flocks of up to 11 pigeons, using the orthogonal method, and the dynamics of turns was unveiled for the first time. Ikawa et al. (1994) studied swarms of up to 20 mosquitoes in the field, and recorded the nearest neighbour distance.

The time span over which these empirical investigations occurred, and the small numbers of animals considered, illustrate the difficulty of obtaining high-quality 3D data. Even the most recent studies, such as that by Budgey (1998) could only reconstruct the positions of a maximum of 61 lapwings, basically using the same technology employed by Major and Dill (1978) twenty years before. Compared to the advances in most experimental fields of science, the situation in 3D studies of collective animal behaviour is somewhat disappointing. The number of animals in all these studies is very low compared to natural conditions, where groups can range up to thousands, or even hundreds of thousands of individuals. As we note in our companion paper (Cavagna et al. 2008), analysing small groups has some serious drawbacks, related to the preponderance of border effects. Moreover, compactness of the groups in all these analysed cases has been quite poor, due to the technical difficulties of dealing with packed groups of animals.

Notwithstanding these criticisms, it must be noted that the first generation of empirical studies (1970-2000) was essential to establish the field of collective animal behaviour on a firm basis and their value should not be underestimated. They gave a first realistic perception of the structure of animal groups and their dynamic properties. Moreover, the experimental techniques were sound. Although there was a need for them be refined and developed, in order to produce more substantial results, the evidence suggests that the direction taken by these studies was the right one.

However, in the last twenty years the main focus of the community has shifted to a different target, leading to a drastic reduction in the number of empirical investigations and of 3D field studies in particular. With the advent of large computers, the field of collective behaviour has been dominated by numerical models (Sakai 1973; Aoki 1982; Reynolds 1987; Heppner & Grenander 1990; Huth & Wissel 1992; Vicsek et al. 1995; Gregoire & Chate 2004). Of course, models are extremely useful. The fundamental ingredients of a model can be kept under complete control, so that it is possible to understand clearly what is the connection between a certain phenomenon (the model's output) and the biological ingredients causing it (the model's input). It is mainly thanks to models that we now understand how collective behaviour may emerge as the consequence of self-organization, without the need for centralized co-ordination of individuals.

While models can complement empirical investigation, they cannot replace it. Apart from some very general issues that can be assessed on the basis of purely qualitative observations (as, for example, the emergence or not of collective behaviour), a model's outcome should always be compared to empirical data. Without this essential 'ground-truthing', models can proliferate in an uncontrolled fashion and, as a by-product, so does the number of different theoretical frameworks. Selecting the right model, and deciding the correct hypothesis, becomes merely a matter of taste, or guesswork, in absence of real data.

As feared by Davis (1980), we believe that, within the field of collective animal behaviour, speculation has outgrown empirical groundwork. As we show in this paper, new tools from statistical physics and computer vision can now help solving the major outstanding problems of 3D reconstruction, and so allow a much needed return to empirical data collection. Using such tools, we managed to reconstruct, for the very first time, the 3D positions of several thousand individual birds in cohesive flocks under natural conditions (Ballerini et al. 2008a, b). Compared to previous empirical studies, which considered few tens of loosely organized animals, this is an advance of two orders of magnitude in the empirical study of collective animal behaviour. This result has been achieved in the context of STARFLAG (*Starlings in flight: understanding the patterns of animal group movement*), a project financed by the European Commission in the context of the 6$^{th}$ framework programme. Within STARFLAG, biologists, physicists, and computer scientists joined forces to study collective animal behaviour, both with an empirical and with a theoretical approach. We report here the main methods developed by STARFLAG to obtain empirical 3D data of animal groups.

The key to achieving the STARFLAG results was the solution of the correspondence, or matching, problem. This has been the most serious bottleneck of all former empirical studies. When using any 3D technique, one must place into correspondence different images of the same animal. For example, in stereometry, one has two images of the group taken from two different points of view. In order to perform a 3D reconstruction, one must take a given animal's image on one photograph, and identify the corresponding image on the other photograph (Osborn 1997). The matching problem is particularly severe when there are many similar animals positioned very close to each other, which, unfortunately, is typical in natural flocks. Until now, no computer algorithm has been able to do this automatically, and thus the matching was performed by hand. Clearly, this severely limited the number of animals and the density of the groups that could be studied. STARFLAG solved this problem by using a blend of statistical physics, computer vision and mathematics.

Some of the techniques used by STARFLAG may be unfamiliar to scientists working in the field of animal behaviour. In this paper, we explain all these techniques and provide the necessary tools to reproduce our results, illustrating our points using the 3D empirical study of STARFLAG. The paper is divided into two main parts. First, we will explain how to set up an apparatus for the 3D reconstruction of large, cohesive and potentially distant animal groups. Our method is stereometry, so we will also give a very brief reminder of this technique. The second part of the paper explains how to transform a set of stereoscopic images into a list of individual 3D coordinates of the animals within the group. Here, our solution to the correspondence problem plays a major role. This part of the paper is, therefore, the most technical, as it was heavily inspired by statistical physics, but it is also the most innovative part of the entire work. Details of how to analyse 3D data sets are given in our companion paper (Cavagna et al. 2008) where the techniques described can be applied to the results of numerical models, in addition to empirical data.

## SETTING UP THE APPARATUS

We begin this section with a discussion of digital imaging; in particular, the problems that arise when working with commercial, non-metric cameras. We then introduce some basic notions of stereometry, and show how a careful analysis of the required experimental accuracy fixes most of the technical parameters of the apparatus, such as the distance between the cameras and the necessary alignment. Finally, we deal with the synchronization problems of standard commercial cameras, and how these can affect measurement accuracy.

The 3D reconstruction of animal groups can be performed at two levels the global and the individual. At the global level, one describes the group as a whole, disregarding the 3D position of the individuals. For example, with sonar techniques, it is possible to study a school of sardines giving its overall position, shape, size, and even the 3D map of its inner density, without knowing the position of individual fish. At the individual level, in contrast, the 3D positions of all (or as many as possible) animals in the group are calculated. Given a set of individual 3D positions, it is possible to reconstruct all global quantities, whereas global descriptions obviously miss the information related to individual behaviour. A global description is thus unable to reveal how collective behaviour emerges from individual interactions, perhaps the most fascinating question in the field. For this reason, 3D reconstruction at the individual level is more valuable than a global level analysis, even though it is technically more difficult. In the present work, we deal only with individual 3D reconstruction.

Cameras

In order to perform 3D reconstruction at the individual level, optical techniques are by far the most precise, compared to radar or sonar. Unless one has the resources to build one's own optical equipment, tailored to the specific needs of the study, most researchers will have to rely on commercial digital still or video cameras. Cameras are normally the most expensive piece of equipment in the apparatus, so that the entire budget for a study can be gauged on their cost. It is therefore seems wise to start this handbook with a brief discussion of camera-related problem (Adams 1983).

Still Cameras vs. Video Cameras: a Matter of Resolution

If one is only interested in studying the static structure of a group then, of course, multiple time images are not needed. On the other hand, if one wishes to dynamically track individual animals, the very first question to ask is: still cameras or video cameras? Typically, still cameras have much higher digital (i.e. pixel) resolution, but rather low refresh rate, i.e. a small number of frames-per-second (fps), while video cameras have high refresh rate, but lower digital resolution. It is therefore important to consider the trade off between these two factors. We start by discussing digital resolution.

There are two very important requirements to meet in this respect: first, we need to resolve different nearby animals on the image; second, the whole group must be entirely contained within the field of view of the camera, i.e. the image must be entirely contained in the camera's charge-coupled device (CCD). This second requirement follows from the fact that we operate with immobile cameras, which do not track the group (alignment issues become severe with moving cameras). We call $E$ the width in meters of the group, $e$ its width in pixels on the CCD, $R_0$ the minimal inter-individual distance in meters, and $r_0$ the same distance in pixels on the CCD. In order to satisfy the first condition, it is reasonable to require that different animals are separated, on average, by at least 4 pixels, and therefore to set $r_0 > 4$. For the second condition, we require that the optical dimension of the group must not exceed, on average, 1/3 of the length of the CCD, otherwise the group is not going to remain in the field of view for very long. We therefore require that $e < n/3$, where $n$ is the total number of pixels of the CCD along its shortest direction. These two basic requirements therefore imply:

$$\frac{r_0}{e} > \frac{12}{n} \quad . \quad (1)$$

Simple proportions fix the ratio between a length $L$ in real space and its projection $l$ onto the CCD (Fig.1). We have:

$$\frac{l}{L} = \frac{f}{z+f} \quad , \quad (2)$$

where $f$ is the focal length and $z$ is the distance of the target, both measured in meters. We note that typically $f$ is much smaller than $z$ (for example, 35mm vs.100m), so that without loss of generality we can write:

$$\frac{l}{L} = \frac{f}{z} \quad . \quad (3)$$

This equation implies that the ratio between projected length and real space length is the same for all targets at the same distance $z$, and therefore:

$$\frac{e}{E} = \frac{r_0}{R_0} \quad , \quad (4)$$

Putting together relations (1) and (4), we obtain the minimum amount of linear pixels $n$ needed to meet both our requirements:

$$n > \frac{12 E}{R_0} \quad . \quad (5)$$

A typical example from the STARFLAG project is a flock of starlings with a width $E = 40$m and an inter-individual distance $R_0 = 0.2$m. In this case we need about 12x40/0.2=2400 pixels in the shortest frame direction, which means a total of 3600x2400=8.6 Mega pixels (Mp) in a standard 3:2 image ratio. Note that this example is quite typical of many kinds of large and cohesive animal groups: the ratio $E/R_0$ is an indication of the linear number of animals in the group, so it cannot be very small for groups with a large number of animals.

At the time of writing, standard commercial video cameras are far from this level of digital resolution: even high definition (HD) video cameras only have around 1.0Mp. Indeed, very few products at the moment have significantly higher resolution, and they are very expensive. Still digital cameras, on the other hand, have excellent pixel resolution, at the price of slower frame rate, typically around 5fps (higher frame rates cause desynch problems –see below). This is too small for tracking most kind of animals in 3D. However, we will describe later how the refresh rate of the apparatus can be improved by doubling (or more) the number of cameras. The final message is that, in order to study cohesive and large groups, high-resolution digital cameras are probably the best choice at the moment.

Group Distance and Focal Length

Relation (5) between number of pixels $n$, width $E$ and inter-individual distance $R_0$ does not depend on the distance $z$ of the group. When this distance increases, the magnification decreases, but in principle one can compensate for this by choosing a lens with a longer focal length, thus keeping magnification constant. In fact, the relation between focal length $f$ and minimal inter-individual distance $R_0$ can be derived from equation (3),

$$\frac{r_0}{R_0} = \frac{f}{z} \quad . \quad (6)$$

We have to be careful here, however: if we measure the inter-individual distance $r_0$ on the CCD in pixels, we must do the same for the focal length. We call $\Omega$ the focal length in pixels, such that $f = \Omega p$, where $p$ is the linear length of a pixel. We can now write:

$$\frac{r_0}{R_0} = \frac{\Omega}{z} \quad , \quad (7)$$

where now both $r_0$ and $\Omega$ are measured in units of pixels. If we require, as we did above, that the inter-individual distance on the CCD must be at least 4 pixels, $r_0 > 4$, we get:

$$\Omega > \frac{4z}{R_0} \quad . \quad (8)$$

For example, for a group at $z=150$m, with an inter-individual distance $R_0=0.2$m, equation (8) says that we need a lens with focal length $\Omega > 3000$ pixels. Considering that the linear size of a pixel in standard digital cameras is $p \sim 8.2 \times 10^{-3}$mm, we get a minimal focal length $f =24$mm. If the distance $z$ is now doubled to 300m, we have to double the focal length $f$ in order to keep the image's magnification the same. If we do not do this, the image becomes too small, and the inter-individual distance $r_0$ on the CCD falls below the desired number of pixels (~4). In a typical STARFLAG photograph, the linear size of a bird at about 100m is between 5 and 15 pixels.

Therefore, from equation (8), one can conclude that, in order to consider groups at longer distances, one can, in principle, simply use lenses with longer focal lengths. In practice, this is not the case. The maximum distance, $z$, is sharply limited by the fact that the optical resolution of a system decreases very steeply with distance from the target. This is mainly due to the diffraction from atmospheric mist, especially in urban areas, which increases drastically with distance from the target. On the other hand, using very short focal lengths also has its shortcomings, because these lenses normally have substantial optical distortion and a larger number of internal elements. It is advisable, therefore, to use 'normal' lenses (i.e. lenses whose focal length is comparable to the CCD size), since they offer a good balance in terms of resolution, distance and dimension of the groups.

Within STARFLAG, we used Canon EOS 1D Mark II cameras, whose CCD size is 28x19mm, corresponding to 3504x2336=8.2Mp. We mounted Canon lenses with focal length $f=35$mm lenses (maximum aperture f2.0), corresponding to $\Omega =4273$. This allowed us to have to have a very good magnification, at the cost of a slightly smaller field of view.

Camera Calibration

The most serious drawback in working with commercial (albeit high end) cameras is that they are not precision instruments. This means that they are not conceived to produce a quantitative measure, but rather a qualitative representation of reality. As a result, the nominal specifications of each camera may differ substantially from the real specifications. If not properly accounted for, this discrepancy can have an enormous impact on accuracy of measurements.

The most basic model of a camera is the pinhole model (Hartley & Zisserman 2003). This model is specified by the focal length $\Omega$ and by the *x-y* position $u$ and $v$ of the optical axis with respect to the CCD centre (all these quantities are expressed in units of pixels). Knowing these values is crucial, since they are an input of the 3D reconstruction formulas. Therefore, they have to be very precisely determined. Moreover, all lenses are affected by radial distortion (Forsyth & Ponce 2002), which is particularly large close to the borders of the field of view. Such distortion may cause a shift of an image point of up 20-30 pixels (see Figure 2), with catastrophic effects on the final accuracy of the 3D apparatus.

The only way to transform a commercial camera into a precision instrument is to have each camera professionally calibrated. In this way, the focal length $\Omega$, and the coordinates $u$ and $v$ of the optical axis are carefully measured (this should be done at a temperature close to the final working temperature). For example, the nominal value of the focal length of the lenses we used within STARFLAG was 35mm, equivalent to $\Omega$=4273 pixels. The average calibrated value of the focal lenght, on the other hand, was $\Omega$=4335 pixels (different STARFLAG cameras had a calibrated focal length that differed one from another of up to 50 pixels).

Professional calibration also includes the measurement of the four parameters of radial and tangential distortion within a quadratic expansion (Forsyth & Ponce 2002; Hartley & Zisserman 2003), such that each image can be at least partially corrected for distortion. Note that it is quite difficult to perform camera calibration in an unspecialized laboratory. In particular, we had unsatisfying results with ready-to-use calibration routines that rely on 2D targets grids.

Camera calibration has enormous importance. At the same time, professional camera calibration is not a panacea: some radial distortion may persist even after this procedure. Tests in reconstructing the known 3D positions of targets, like those we describe later, must always be performed to check the accuracy of the apparatus. Finally, we note that camera calibration must be performed at least annually.

Focus, Contrast and Exposure

Sharpness is a key issue, especially in dense and cohesive groups. For this reason, focus must be carefully controlled. Cameras must be switched to manual focus, since auto focus is slow and inaccurate. Given that, typically, animals are quite distant from the cameras, the most common error is to set the camera's focusing ring to infinity. In principle, this is the correct choice: for example, with a focal length of 35mm, targets beyond 50m are, for all practical purposes, at infinity. However, when the focusing ring is set to the default infinity mark, a standard camera is in fact focusing beyond infinity, and photos will be out of focus significantly (see Fig.3). The reason why camera manufacturers do this is to compensate for the fact that, at different temperatures, the materials in the lens expand and contract, which changes the infinity focus-point. Allowing the lens to focus to what, in average situations, is beyond infinity will, under extreme temperatures, just allows the lens to focus at infinity. Therefore, the focus of each camera used must be calibrated manually before field observations begin. This is achieved in the following way: by using a standard laser meter, one can locate targets at about the same distance as the groups of interest, and take several test photos at various positions of the focusing ring. One can then find the optimal focus position, and mark on this on the focusing ring such position. An important remark is in order: when the camera gets calibrated it is important to specify what is the distance at which it will be focused, since the focal length is the distance between focal plane and rear nodal plane.

Sharpness is also significantly affected by aperture, shutter speed and sensitivity (ISO). First of all, aperture and shutter speed must be set manually. This is essential for the camera to save time, and thus have a faster shooting rate and a more accurate synchronization with other cameras (see later). Ideally, in good light conditions, maximum sharpness is obtained with low ISO (which implies low noise), slow shutter speed, and the aperture set in the mid-range (to increase depth of field). There are, however, several constraints, mainly set by the need of a decent exposure and, in particular, by the need to obtain a reasonable contrast between animals and the background. As we shall see later, in order to automatically recognize animals, one uses a luminosity threshold, in order to separate them from the background. For this to be effective, it is important to maximize the difference in luminosity between animals and background. The best way to achieve this is to have a well-balanced exposure, meaning that the average level of the image's exposure should be in the mid-range. In strict photographic terms, this is equivalent to saying that, when using a spot light meter, the background (for example, the sky) should be placed as close as possible to zone V, i.e., to 18% grey (Adams 1983). This prevents the contrast-flattening effect of underexposed and overexposed images. Achieving perfect exposure can be tricky, especially if photos have to be taken at dusk (as with starling flocks). In this case, one must find a reasonable trade off between ISO and shutter speed, to optimize sharpness and contrast.

Determination of optimum aperture is tricky: a small aperture increases depth of field, and thus enhance sharpness, but even just a few stops below full aperture can decrease drastically the camera's refresh rate, and thus compromise camera synchronization. This is due to an associated increase in release time lag (see below). In the STARFLAG project, we found that a maximum of two stops below full aperture granted a reasonable depth of field, together with accurate synchronization (full aperture of our lens was f2.0, so we used down to f4.0). In some cases (i.e., dusk or dawn), light conditions change rapidly, so that exposure settings must be changed during the course of observations.

Elements of Stereometry

The best technique to reconstruct the 3D position of distant animals in the field is probably stereometry (Longuet-Higgins 1981). Compared to other techniques, most prominently the orthogonal method, stereometry is more flexible, in that the group's distance is limited only by optical resolution, and not by being enclosed within the apparatus. Stereometry was the technique used by the STARFLAG project, and we provide a brief review of its most basic elements.

The fundamental idea of stereoscopy is that, by taking images of the same object from two different positions in space, we have enough information to reconstruct its 3D coordinates, provided that we know the mutual position of the two cameras. The mathematics behind this is simple: the two images consist of a series of two pairs of 2D coordinates, and these four values are sufficient to work out the three independent 3D coordinates. Even though the principle is simple, the actual equations are quite complicated, and it is essential to refer to the most recent technical literature (Hartley & Zisserman 2003) in order not to make gross mistakes. Here, we sketch the main equations in their simplest form. This is sufficient to conduct an error analysis, and thus determine the key parameters of the apparatus.

The Minimal Baseline

In Fig.4 we depict the simplest stereoscopic setup: two parallel cameras are separated by a distance $d$ in the $x$ direction. This separation is the baseline, and the first experimental question we have to answer is: how large must $d$ be in order to achieve sufficient sensitivity in the apparatus? Clearly, the larger the baseline, the higher the sensitivity (indeed, with $d=0$ there would be no sensitivity at all!). To quantify this, we write the fundamental equation of stereoscopy:

$$s = u_L - u_R = \frac{\Omega d}{z} \quad . \quad (9)$$

In this expression $s$ is the stereoscopic shift, or disparity, namely the separation in pixels between the two images. This is the difference between the abscissa of the image in the left photo, $u_L$, and that in the right photo, $u_R$. $\Omega$ is the focal length expressed in units of pixels, $d$ is the baseline in meters, and $z$ is the distance of the object in meters. We stress that this equation only holds when the optic axes of the two cameras are perfectly parallel, when the separation $d$ is purely along the $x$ direction, when the calibration parameters are exactly the same for the two cameras, and when there is no radial distortion. In this case, the two images are shifted purely in the $x$ direction, while they have the same photographic ordinate $y$, which goes through the plane in Fig.4.

Imagine that an animal initially at distance $z_1$ moves to a distance $z_2 = z_1 + \delta z$. This will cause a change in disparity, $\delta s$. The relation between $\delta s$ and $\delta z$ can be computed by differentiating equation (9):

$$\delta z = \frac{z^2}{\Omega d} \delta s \quad . \quad (10)$$

In order for our apparatus to be sensitive to this shift, it is essential that the variation, $\delta s$, of the disparity is larger than one pixel; otherwise the space separation, $\delta z$, will be undetectable. We can also invert the argument: it is not possible to locate the position of an animal on the image with a resolution better than one pixel, so that there is an unavoidable error, $\delta z$. Therefore, the minimal baseline necessary for resolving the shift $\delta z$ is obtained by setting $\delta s > 1$:

$$d > \frac{z^2}{\Omega \, \delta z} \quad . \quad (11)$$

This is a very instructive equation, and one of the most important in stereo analysis: the minimal baseline grows very fast with the distance of the target, and, as expected, it is larger the lower $\delta z$, i.e. the higher the required spatial sensitivity. For example, for a standard 35mm lens, we have $\Omega = 4273$. If we need to resolve birds at a distance $z=150$m, with a sensitivity $\delta z=0.5$m, we get $d > 10$m. Note that 0.5m is by no means an exaggerated sensitivity, considering that birds may get as close to each other as 0.3-0.2m. If we want to reconstruct a group 200m away with the same accuracy, we would then need $d > 20$m. Most previous stereo studies have tried to reconstruct animals (typically birds) at similar distances from the apparatus, but with baselines in the 2m-5m range: inverting equation (11) shows that this amounts to a resolution $\delta z$ between 2.3m and 4.6m; hardly satisfying if we want to be able to say something accurate about the relative distances among individuals. Within the STARFLAG project, we used a baseline of $d=25$m, sufficient to have a decent accuracy for starling flocks at a distance of up to 200m.

The Need for Alignment

As we have seen, the error in the relative distance between animals is dominated by the pixel resolution, $\delta s$, which cannot be smaller than one pixel. There is, however, a second important source of experimental error, which comes from the misalignment of the apparatus. If the two cameras are non-parallel, and have a convergence (divergence) angle α between them, equation (9) is modified:

$$s = \frac{\Omega d}{z} - \Omega \alpha \quad , \quad (12)$$

(this equation is only valid for small values of α). An error, δα, in the determination of the angle, produces an error, Δz, in the absolute determination of the distance, z:

$$\Delta z = \frac{z^2}{d} \delta \alpha \quad . \quad (13)$$

This equation tells us that the smaller the baseline, $d$, the smaller the alignment error, δα, must be in order to keep the error, Δz, as small as possible. The situation looks quite grim: even with a 20m baseline, in order to make an error, Δz, smaller than 1m in the determination of targets at $z$=100m (an accuracy of 1%), one would need to align the cameras with an accuracy better than 0.002rad (or 0.11°). The same 1% accuracy on targets 300m distant would require an alignment better than 0.0006rad!

    Any misalignment in the cameras affects absolute distances more than relative ones, meaning that the positions of two animals close to each other in space will be shifted by a similar amount, with little effect on their inter-individual distance. However, this is small consolation: inter-individual distances may not change substantially, but the shape of the group as whole will be altered considerably, with serious consequences for the calculation of all group morphological properties.

    Of course, the situation improves dramatically if the average distance of the group from the apparatus is significantly lower than those considered so far in our discussion. This is typically the case in a laboratory. For example, in order to study a 2m long school of fish in a tank at a 5m distance from the apparatus, with the same 35mm focal length lens, and a relative resolution of 0.05m, a baseline of 1m is more than adequate. Even here, however, an alignment better than 0.002rad is needed, although this is much easier to achieve with two cameras mounted on the same support just 1m apart. When recording measurements in the field, however, it is often very difficult to get close to animal groups without perturbing them, and it is practically impossible with birds.

    A long baseline (>20m) is necessary, therefore, for the sake of accuracy, because both the error on relative distances and the error on absolute distances are inversely proportional to the baseline. A long baseline is also a big inconvenience, however, for two good reasons: first, it is impossible to mount the two stereo cameras on the same support, which makes it very difficult to align them to the desired accuracy (<0.002rad), especially in the field, where many lab facilities are absent. Second, a long baseline implies a large stereometric disparity, making it very difficult to match different animals in the two photos. The latter problem is the hardest one, and we will deal with it in the second part of the paper. For now, we concentrate on camera alignment.

Accurate Alignment in the Field

In the STARFLAG project, we set the baseline to 25m, and we operated on the roof of a public building, where we were not allowed to leave any piece of equipment. This meant that we had to mount and dismount the whole apparatus on every observation session. These constraints forced us to discard any form of optical alignment, such as lasers.

Lasers are very helpful in the lab, where they can be carefully calibrated on a single occasion along with the rest of the apparatus, but much less useful under the uncontrolled conditions typical of animal observations in the field.

When trying to develop an effective alignment method, we also had to consider the following fact: working with parallel cameras is very inconvenient, because the maximum overlap of the two fields of view corresponds to an infinite distance of the targets. If the average distance of the group is $z$, it is optimal to have the two cameras slightly convergent at an angle $\alpha=d/z$. In the STARFLAG setup, we used a convergence angle of $\alpha=0.22$rad (12.6 degrees), giving optimal overlap of the fields of views at about 100m, the average distance of the flocks from the apparatus. Precisely fixing a non-zero angle between two distant cameras is, however, a non-trivial task, and is much more difficult than aligning cameras in a parallel fashion.

Within STARFLAG, we developed a basic, but effective method (Fig.5). Each camera was mounted on a rigid bar, made of an aluminium alloy that deformed as little as possible under temperature changes. The bar was of length $L$ and the camera was aligned to the back of the bar. The bar was then mounted on a sturdy professional tripod, with a micrometric head (we used Manfrotto 475 tripods and Manfrotto 400 heads). A thin line, as inextensible as possible, connected the external sides of the two bars to each other. The line was under strong tension, so that it was very rectilinear (this can be tested with a laser). We used a fishing line with a diameter $a=0.25$mm. On the internal sides of the two bars there were two fine gauges, over which the line passed. Once the line was tense and rectilinear, it was possible to adjust the micrometric head until the line passed exactly over the desired reading, $H$, on the two gauges (Fig.5). This fixed the mutual convergence angle between the two cameras to the value:

$$\alpha = \frac{2H}{L} \quad . \quad (14)$$

The alignment precision is given by the ratio between the bars' length and the reading error of the position of the line over the gauge, which is comparable to the line's diameter. This gives:

$$\delta\alpha = \frac{a}{L} \quad , \quad (15)$$

STARFLAG used bars 680mm long, giving an alignment accuracy of $0.25/680 = 0.0004$rad, well within the required accuracy of most stereoscopic observations. The only problem with this alignment method is wind, since it moves the line, making it hard to align the line to the gauges. However, as long as the wind is below 12m/s, the method works well, with a baseline of up to 25m.

The convergence angle $\alpha$ is only one of the three Euler angle that specifies the mutual orientation of the two cameras. Using flight terminology, $\alpha$ is the yaw angle, corresponding to a rotation around the $y$-axis. The roll angle $\gamma$ corresponds to a rotation around the $z$-axis, and finally the pitch $\beta$ corresponds to a rotation around the $x$-axis (Fig.6). In the STARFLAG setup, the mutual roll angle $\gamma$ between the cameras was set to zero. This can be done very accurately using the line method above: on each bar, we mounted a second gauge parallel to the $y$ axis and rotated the micrometric heads until the line passed over the desired marks on both bars (Fig.5).

In contrast, the mutual pitch $\beta$ cannot be fixed using the line. In STARFLAG, we used a Suunto clinometer to measure the tilt-up of the two bars (typically around 30-40%) and thus to set to zero the mutual pitch angle. The maximal error of the clinometer was 0.02rad, which, by default, became the error on the relative pitch. This is much worse than the error on the yaw and roll angles (~0.0004rad). Fortunately, however,, the error on the pitch angle $\beta$ has very small influence on the accuracy of the 3D reconstruction formulas (Hartley & Zisserman 2003). Various tests performed within STARFLAG (see later) showed that an error on the relative pitch of 0.02rad is more than adequate to achieve very accurate results.

The last thing that must be measured is the baseline, i.e. the distance $d$ between the two cameras. This can be done accurately and easily using a commercial laser meter. One point that deserves attention, however, is that, strictly speaking, the two cameras are separated by a 3D vector, and not simply by a scalar quantity (Hartley & Zisserman 2003). Let us take as a reference frame the one of the right camera, with the $x$ and $z$-axes as in Fig.4. If the cameras are parallel (and so are their focal planes), then the $x$-axis runs parallel to both their focal planes, so that all the displacement, $d$, takes place along the $x$-axis. Under these conditions, the displacement vector of the left camera has components ($-d$, 0, 0). If, on the other hand, the cameras are convergent by an angle $\alpha$ (yaw), then the displacement vector of the left camera has a small $z$ component, and it reads ($-d$, 0, $d\alpha/2$), where we have neglected second order terms in $\alpha$. We repeat once again that an accurate determination of the alignment angles, and of the baseline length, is of paramount importance to generating reliable results, because all these parameters enter critically in the reconstruction formulas.

Stereometric Tests and STARFLAG accuracy

The 3D apparatus must be carefully and regularly tested. This can be achieved by taking a set of stereo photos of some artificial targets located at a distance comparable to the animal groups under study (the optimal stage is the façade of a building; targets can be located at different windows). The mutual distances among the targets and their absolute distance from the origin of the reference frame (normally the focal point of the right camera) are then measured by means of a Hilti laser meter. The laser estimate is then compared to the 3D streometric reconstruction, so that an assessment of error can be made. It is important to have several targets, covering as many different parts of the field of view as possible, since radial distortion strongly depends on the position on the field of view. Tests must be performed regularly, in order to be sure that the calibration parameters are not drifting away from their original values. Note that, in order to work out the 3D coordinates of the targets, one needs to have the reconstruction algorithms ready. The full reconstruction formulas are more complicated than the simplified formulas above, and they will be described at the end of the second part of this paper.

Within the STARFLAG project, we performed several stereometric tests. The baseline was $d$=25m. The focal length in pixels of our lenses (Canon EOS 1D Mark II with 35mm lenses) was $\Omega$=4335 (this is the average value, since the calibrated value varied from camera to camera of about 10 pixels). The error, $\delta z$, on the relative distance of two nearby targets located at distance, $z$, from the cameras is dominated by the error. $\delta s$, in the determination of the image positions. For birds at a distance, $z$=100m, the equations above give a nominal error on the relative distance equal to $\delta z$=0.09m, for two objects at 1m one from another. The error, $\Delta z$, on the absolute distance is dominated by the error $\delta\alpha$ on the convergence. With bars 680mm-long and a line of diameter 0.25mm, our method gave a maximum misalignment $\delta\alpha=0.25/680=3.7\times10^{-4}$rad, thus giving a nominal error $\Delta z=0.14$m (targets at 100m). Our tests gave $\delta z<0.04$m, corresponding to $\delta s<0.4$pixel, and $\Delta z<0.92$m, corresponding to $\delta\alpha<2.3\times10^{-3}$rad. The maximal error on the absolute distance was thus small, but larger than the nominal value expected from our alignment method. A careful investigation of this fact showed that residual radial distortion was responsible for this greater-than-nominal error.

Synchronization

Whenever the targets are on the move, which is the case for most animal groups, stereo photographs must be taken at the same instant. This is done by connecting both cameras to an electronic device that shoots them synchronously. Of course, perfect synchronization is not possible. A certain amount of desynchronization is always present and one must carefully check accuracy.

The Release Time Lag

The largest source of synch error is the cameras themselves: in particular, fluctuations in the release time lag (RTL). The RTL is the lapse of time between the moment the camera is fired and the moment the shutter curtain actually moves, allowing light to reach the CCD. Each camera model has a nominal RTL, but in practice this fluctuates considerably. Clearly, a different RTL in the two stereo cameras can be a major cause of desynchronization (desynch). The best means of stabilizing the RTL is to keep the lens' aperture as large as possible. For most cameras, the RTL is stable if the aperture is kept within two stops from the largest value allowed by the lens. The reason for this is that, of the many things contributing to the RTL, one of the most prominent is the time taken by the camera to close down the aperture. Consequently, the smaller the aperture, the longer and more erratic the RTL becomes. Note that this remains true even when the exposure is preset manually.

The effects of desynch are easy to calculate if the animals' typical velocity is known. The distance travelled by animals within a desynch time, $\delta t$, is simply $v \times \delta t$, where $v$ is the velocity. The effect of desynch is therefore to create a false extra stereoscopic shift, which is solely due to animal movement, rather than to *bona fide* distance. This false shift will therefore alter the reconstructed 3D position of the target. A reasonable rule of thumb is that this desynch distance should be smaller that the physical size of the animal. In the STARFLAG setup, we measured the desynch to be smaller than 10ms, which for birds travelling at about 10m/s gives a desynch distance of 0.10m, acceptable when compared to starlings' wingspan of 0.40m. The RTL of a commercial camera should never be trusted, and must always be measured carefully.

Shooting Multiple Images

So far we have explained how to synchronize two cameras on a single photo. This is fine if one is interested only in the static structure of groups. However, much of the interest in the 3D reconstruction of animal groups lies in determining their dynamical properties, and to achieve this aim, one needs to take several successive images, at the fastest possible rate. Excluding video cameras, whose shortcomings we have already discussed, this means firing two cameras at a fast rate, while keeping them closely synchronized. As with synchronization for single photos, this is a non-trivial task.

High-end cameras have a continuous mode: by keeping the shutter release button pressed down, the camera shoots continuously at a certain rate, measured in frames-per-seconds (fps). Although intuitively attractive, a camera's continuous mode must not be used for the purposes of dynamic 3D reconstruction. This is because a camera whose nominal refresh rate is, for example, 8fps, does indeed shoot eight photos in a second, but it does not shoot one photo every 1/8 of a second. Shooting during continuous mode is typically quite irregular, and fluctuations are large, normally exceeding the minimal threshold for a decent synchronization.

The only solution to this problem is to fire each individual camera independently, using a purpose-built electronic device. Even so, one needs to be very careful: the maximum shooting rate must be kept well below the nominal continuous rate, otherwise the RTL will fluctuate too much, and synchronization will be lost. For example, in STARFLAG, our cameras had a nominal continuous rate of 8.5fps. After accurate tests, however, we discovered that, if the camera was fired faster than 5fps, the RTL became erratic, with fluctuations that were too large to guarantee reasonable synchronization. Thus, we had to fire the cameras at 5fps, rather than at 8.5fps. The only way to determine the maximum shooting rate that a camera can sustain without becoming erratic is to conduct appropriate tests, such as taking photos of a high precision chronometer. In the STARFLAG project, we built a simple LED-based device that displayed the time in milliseconds and then we photographed it to check the cameras' synchronization.

For birds that fly fast as starlings, a rate of 5fps is not high enough to produce satisfactory dynamic tracking of the animals' trajectories. Even though this problem is likely to be solved in the near future by improving cameras technology, it is worth briefly discussing this issue. Our solution to this problem was to use twin pairs of cameras shooting in an interlaced fashion (Fig.7). That is, two cameras (1 and 2) were mounted as close as possible to each other on the bar. The electronic device fires both cameras at 5fps, but with opposite phases, so that the net effect is a refresh rate of 10fps of the twin camera system. The two cameras must be carefully aligned, but this is relatively easy, since they are mounted rigidly on the same bar.

This solution comes at a cost, however. First, doubling the number of cameras means doubling the budget of a study. As we discuss below, our matching procedure requires a "trifocal" method, with three points of view this means a total of six cameras is needed. Second, interlacing has a small but disturbing effect on 3D reconstruction: the two series of photos are taken from two coordinate frames (1 and 2) that are slightly shifted (0.2m or so) with respect to each other. In addition, the two stereo couples (1Left-1Right vs. 2Left-2Right) will always be slightly different from each other, because it is impossible to achieve identical calibration and alignment for the two stereo couples. The shift of the reference frame is easy to correct. However, the difference in calibration and alignment is more annoying: the dynamic trajectory of a target is an interlaced series of reconstructions, 1-2-1-2-1-etc. Given that systems 1 and 2 reconstruct in a slightly different way, the resulting 3D trajectory therefore resembles a zigzag path. The magnitude of this effect depends on the specific apparatus used. Within STARFLAG, the effect was small, in the order of a centimetre. At one level, this low value represents a good consistency check for the whole 3D reconstruction procedure and the reliability of the error analysis. However, even a small value like this quickly becomes very annoying as soon as one begins to look at dynamic trajectories. In this case, even small variations of the heading of a bird can make big differences when computing individual velocity. Compensating for this zigzag effect is possible, although not completely straightforward: once a certain number of trajectories has been determined, one can then calculate the optimal transformation (rotation plus translation) of coordinate frame 2 that minimizes the overall roughness of the trajectory. We will discuss this point further in a separate publication on individual dynamic trajectories.

# FROM 2D IMAGES TO 3D POSITIONS

Once digital stereoscopic images of an animal group have been taken, there is still some work needed to generate a 3D reconstruction. First of all, individual animals on the photos must be recognized and their 2D, photographic positions measured carefully (Fig.8). This process is technically known as *segmentation*. This is a relatively straightforward process if animals' images are large and well separated from each other (unlikely case), or relatively hard when animals' images are small, packed and partly overlapping (likely case). After the segmentation, one deals solely with the sets of 2D photographic positions. This brings us to the hardest part of the task, namely stereoscopic matching, also known as the correspondence problem. Basically, we have to match each animal's image in the right photo to its corresponding image in the left photo. This is done for thousands of very tightly packed identical points, because animals' features are of no use, mainly due to distance, poor light conditions, and stereo disparity (Fig.9). Solving the matching problem represents the major original advance of the STARFLAG project. Finally, once the matching is done, the 3D position of each animal can be worked out by using standard algorithms. The key references for this section are Forsyth & Ponce (2002) and Hartley & Zisserman (2003).

## Segmentation

The basic principle behind segmentation is quite simple: animals are dark objects against a lighter background (or, more rarely, vice-versa), so that a threshold on the intensity of the luminosity signal can distinguish between animal and background in the photograph. A simple code can then be run to identify all connected clusters of pixels with intensity lower than the threshold. Each cluster corresponds, in the best-case scenario, to a single animal. In this case, the centre of mass (i.e. the average position of all pixels in the cluster) is calculated, and this gives the best estimate of the 2D position of that animal. In practice, the problem is more complicated than this.

Background Subtraction

The first, and most prominent, problem is that there are many inanimate objects in the photo (trees, buildings, and in particular clouds) that are similar to animals with respect to contrast with the background. Therefore, a naïve, threshold-based method will fail. Fortunately, unlike most unwanted objects, animals move, and this simple fact helps solve the problem. For our purposes here, we assume that animals are darker against a light background.

We take several repeated images, at a given refresh rate (for example 5fps). Given a certain photo, $K$, we want to find a method to subtract its background, so that the only signal remaining is the animals'. To this end, we can consider a short series of $m$ successive photos centred on $K$. For example, for $m=3$ we consider photos $K$-1, $K$, $K$+1. Let us focus on a certain pixel with coordinates $(i,j)$: either this pixel is occupied by the immobile background in all $m$ photos, or it is occupied by a moving animal in at least one of the $m$ photos. If $m$ is large enough, it is very unlikely that pixel $(i,j)$ will be occupied by moving animals in all $m$ photos of the series, because animals are moving. This means that pixel $(i,j)$ has 'seen' the background in at least one photo. We then choose, for each pixel, its highest intensity value within the $m$ photos. Since animals are darker than the background, by doing this we are selecting an instant of time in which that pixel is occupied by the background. In this way, we can reconstruct an image of all immobile objects in the scene (the background), while cutting out all moving animals. (Note that this procedure generalizes the simpler method of subtracting two successive images, and is more reliable)

Once the background is known, we can subtract it from the central photo $K$, and take the absolute value of the intensity difference in each pixel. We thus obtain a new image where only mobile targets, the animals, stand out against a dark background with zero intensity. At this point the threshold algorithm described above can be more safely applied. We finally repeat the same procedure for all photos.

There are, as usual, some important caveats. The first is noise. We normally work with JPG images, at various degrees of compression (in order to minimize the size of the photos and thus optimize the camera buffer), but even in the absence of JPG compression (raw format), some noise is unavoidable. This implies that even after subtraction, the background's intensity will be non-zero. If an animal is very distant or very small, its image will be faint, and its intensity may be barely larger than the subtracted background. For this reason the intensity threshold must be selected very carefully. In order to improve the animal/background contrast it is important to know the colour channel in which the method works best. This strongly depends on light conditions, so that only empirical tests can select the right colour channel. Moreover, after subtraction, it is also useful to apply a noise reduction filter, as median or mean; one must pay careful attention, however, in order to avoid too drastic a reduction in the image's sharpness.

An obvious question raised by this method is determining the ideal number $m$ of images used for background subtraction. The probability that a given pixel is not occupied by an animal's image in at least one of the $m$ photographs, grows with $m$. So, taking large values of $m$ is useful. There is a side effect of increasing $m$, however, because the background may contain some inanimate but slowly moving objects, such as clouds (unfortunately, shooting only on clear or uniformly overcast days would reduce too much the amount of collected data). If the number of photos, $m$, is so large that it covers a substantial lapse of time, clouds may have moved more than one pixel. As a consequence, the subtraction algorithm will identify clouds as moving 'animals', and so introduce them in the analysis. For this reason the number of photos for background subtraction must be carefully tuned to the animals' velocity and size. In the STARFLAG project, we employed a value of $m$ between 3 and 5.

Blob-Splitter

Unless the group is exceedingly sparse, which almost never happens in the most interesting cases, there will always be some animals so close to each other on the photo that they will be recognized as a single objects by the threshold algorithm. We call these objects 'blobs' (see Fig.8). A blob-splitting method is therefore needed.

First, one needs to identify which objects are likely to be blobs. This can be achieved by identifying those objects whose size (area) is larger than expected. Fixing such a threshold in the size may not be easy, however, especially if there is high variability in animals' sizes and shapes. Second, a blob-splitting algorithm must be run on each blob, in order to identify individual animals belonging to the blob. The STARFLAG project used a variant of the watershed method (Forsyth & Ponce 2002): the intensity threshold is raised gradually, until the object splits in two (or more) parts, corresponding to individual animals (Fig.8). The centres of mass of these parts are computed to give their 2D positions. In order to be sure that these parts correspond to individual animals, they are then followed separately by further increasing the threshold, until they finally disappear (i.e. the object's luminosity goes fully below the threshold). If, on the other hand, the objects split again, then the new centres of mass are calculated, until one is sure that only individual animals are left.

The effectiveness of the whole segmentation process, and in particular of the blob-splitting algorithm, can effectively tell us whether or not the group under study is too dense to be reconstructed. If, after careful optimization of the parameters, the segmentation produces huge super-blobs of hundreds of animals, then it is very likely that, even after applying the blob-splitter, the animals' positions will be so noisy that it will be very hard to continue with the analysis. In these cases the only thing one can do is try to improve resolution, both digital and optical (more pixels and better lenses). On the other hand, if blobs contain few animals (up to ten, or a few more), then the blob-splitter can produce excellent results.

Matching

Any method for the 3D reconstruction of a scene (stereoscopy, orthogonal method, shadow method) requires placing different images of the same object into correspondence. This is the correspondence, or matching, problem. For small and/or sparse groups this task is not problematic, and it can be done manually for each animal. However, as soon as the number of animals is larger than a few tens, and the group is more cohesive, the time and effort costs of manual matching become prohibitive (Fig.9).

There are standard methods in computer vision to deal with the matching problem, but they mostly rely on objects' features to do the correspondence. The features used may be colour, shape, size, and orientation, or any other attribute useful for recognizing the same object in two images. Unfortunately, these methods are unsuitable for most animal groups. In order to have the whole group within the field of view, the distance at which photos are taken are often so great that the images of individual animals cannot be identified by their features, and those features that are present are typically the same for all animals. In the case of starlings, for example, a flock looks like a set of dark, featureless points (Fig.9), and matching birds by their features is out of question. Moreover, even in those cases where animals' images are larger and more detailed, it is possible that an animal's feature in one photograph may look significantly different from that in the other one, due to the necessarily large stereo disparity. We conclude that features-based matching cannot be used in animal groups reconstruction.

Elements of Epipolar Geometry

Let us name a certain animal Frank, and let us specify A as the right camera/photo and C as the left camera/photo of the stereo pair. The first practical aid in trying to develop a matching algorithm comes from the following principle of projective geometry: given the position of Frank's image on photo A, we can calculate the straight line in photo C along which Frank's image lies (Hartley & Zisserman 2003) (Fig.10). This is called the *epipolar* line. The epipolar line in photo C is the intersection between the CCD plane of camera C, and a second plane defined by the following three points: Frank; Frank's image on C; Frank's image on A (this second plane is called the epipolar plane). As a simple illustration of this principle, consider two identical cameras, perfectly parallel, and shifted only in the *x* direction. In this case, the two images of Frank have identical *y* coordinates, whereas the *x* coordinate is shifted by an amount equal to the stereo disparity. So, given the image on the right photo, with coordinates $(x_R, y_R)$, the corresponding image on the left photo is found along the line specified by the equation $y_L = y_R$. In more general cases, with non-parallel and non-identical cameras, the epipolar line is no longer parallel to the *x*-axis. The principle, however, is still valid (Armangue & Salvi 2003).

The advantage of knowing the epipolar line associated with an animal is obvious. Given Frank's image on A, we want to identify Frank's image on C. By using Frank's image on A we can compute its epipolar line on C, and instead of searching for Frank's image anywhere in C, we can restrict the algorithm to do so only along the epipolar line. This decreases significantly the complexity of the problem. However, due to noise, the matched image will never lie exactly on the epipolar line, but slightly off it, by an amount that depends on the amount of noise present (Fig.10). As a result, in a dense group, there will be many animals in C sufficiently close to the epipolar line to be confused with Frank. For example, in a starling flock of 2000 birds, knowing the epipolar line may reduce the number of possible candidates (depending on the amount of noise) to 100-200 birds, but this is still a lot of birds. How do we find the right match?

Trifocal Method in a Nutshell

To solve this problem we must bring in a third camera, let us call it B. This camera must be located quite close to the right camera A (say 2.5m), and it must be pointing roughly in the same direction as A (Fig.11). It is important to note that the system A-B will not be used as a stereometric pair (i.e., it will not be used for 3D reconstruction). The only purpose of camera B is to help solving the matching problem. For this reason, cameras A and B need not to be aligned. They still need to have similar fields of views, however, so they should roughly be aligned by eye.

Let us assume that we know Frank's location on both A and on B. In other words, we are assuming that we have solved the matching problem for the nearby pair A-B. This is a reasonable assumption, since the pair A-B has a very small baseline; this fact implies that the stereo disparity is small, making the matching quite easy (we explain later how the A-B matching is performed). There are now two epipolar lines running through photo C, namely the line generated by Frank's image A and the line generated by Frank's image B. The crucial point is that, by construction, Frank's image on C (the match we are after) must lie on both epipolar lines, so that the two lines must cross at Frank's position. This means that we no longer have to look for Frank's image along one line, but at the intersection of two lines, i.e. around a point (Fig.12). This is clearly a great advantage because it dramatically decreases the size of the search area. Due to some inevitable noise, there may be a couple of candidates (including Frank) close to the point of intersection. However, the number of candidates is now reduced enormously, even in very dense groups. In fact, it goes down to less than 5 in a standard STARFLAG case. Moreover, it often turns out that Frank is simply the closest animal to the intersection point, so that he is very easily identified. In this way, the matching problem is (in principle) solved.

Even though this description captures the is the spirit of it, the actual trifocal method is somewhat more complex. For mathematical reasons, rather than using the two crossing epipolar lines, A-C and B-C, as in earlier methods (Ito & Ishii 1986; Ikawa et al. 1994), it is much more convenient to use a mathematical object called a trifocal tensor (Hartley 1997; Hartley & Zisserman 2003). The fundamental idea, though, is the same as described above: if Frank's A-B match is known, it is possible to estimate Frank's position in C. An alternative explanation of the trifocal principle is that by using the A-B match, one can generate a very rough estimate of the 3D position of Frank, and thus project it back onto C. The technical details on how to compute and use the trifocal tensor can be found in the literature (Hartley & Zisserman 2003).

Unavoidable complications

There are two key problems to be solved before the trifocal technique can be used. First, the trifocal technique only works if all A-B matches are known, since the method just transfers every A-B match to C. Even though we expect the A-B matching to be relatively easy, due to the small A-B baseline, it is still unclear how this matching is achieved in practical terms. Second, epipolar lines and the trifocal tensor do not come free. In fact, in order to compute these mathematical objects, a sufficiently high number of matches of the A-B-C triangle must already be known. More precisely, not only do we need all the easy A-B matches, but we also need some of the hard A-C matches to compute the trifocal tensor. This presents something of a paradox: we have a very powerful method to match animals in the essential A-C pair, but we need to know at least some of these matches in order to use this method. Why is this?

A simple example may clarify this point. To know the position $x$, at time $t$, of a car moving at constant speed we can use the equation $x = A + B\,t$, provided that we know the values of the parameters $A$ (the initial position) and $B$ (the velocity). Yet, in order to determine $A$ and $B$ we must already know some points of the trajectory, for example, where the car was after 10sec, and after 90sec. If there is no noise, i.e. no error in the positions, two points are enough, since we must determine two parameters. However, in the more general noisy case, the larger the number of points, the more accurate the determination (via a fit) of the parameters $A$ and $B$, and the better our estimate of the unknown position, $x$, at time $t$.

The epipolar and trifocal relations among the objects in A-B-C are equivalent to the relation between $x$ and $t$ in this example. Similarly, even though the trifocal tensor is in principle determined by 27 parameters (that describe the actual geometric relations among the three cameras), because of noise one needs as many A-B-C matches as possible; in practice, this means at least 50. Once these 50 matches are known, one can compute, via a non-linear fit, the trifocal tensor, which is conceptually equivalent to the two epipolar lines (Hartley & Zisserman 2003).

In order to solve both problems (the A-B complete matching and the A-C partial matching), therefore, we need a matching method that works without any a priori knowledge (or a very poor one) of the geometry of the pair of cameras under consideration. With such a method, we can determine all matches of the easy A-B camera pair. Moreover, the same method can be used to determine the required number of difficult matches (>50) involving the A-C camera pair, and thus to initiate the powerful trifocal process described above. Such a method was first developed by STARFLAG, and forms the core of the whole project. The method is called matching zero (M-Zero), because it is the matching algorithm everything else in the matching procedure is based upon.

Matching Zero

Our brain is quite good at matching. When we look at a pair of stereo images, we clearly see that the global arrangement of points is very different in the two photos because of the stereoscopic disparity (Fig.9). However, what we also recognize immediately is that the *mutual* geometric relationships among points that are *close* to each other are only weakly deformed. In other words, our brain is geared to detecting similar patterns in the two photos. Once detected, we determine the matches simply by identifying the vertexes of these two patterns (Fig.13). M-Zero was born with the aim of mimicking this procedure, and it is, very broadly speaking, a pattern recognition algorithm, although one tailored specifically to epipolar geometry.

Let us consider the M nearest neighbours of Frank in real 3D space, $N_1, N_2, \ldots, N_M$. We call $NR_1, NR_2, \ldots, NR_M$ the images of Frank's neighbours on the right (R) photo, and $NL_1, NL_2, \ldots, NL_M$ the images of Frank's neighbours on the left (L) photo. Finally, we call FR and FL the images of Frank on the two stereo photos (Fig.14). The basic idea is as follows: if the neighbours are close to Frank in real space, then their stereoscopic shift is similar to Frank's. As a consequence, the 2D pattern formed by Frank and his M neighbours must be similar on the two photos. The M-Zero algorithm looks for these similar patterns.

In order to assess the similarity of two patterns, the first thing to do is to superimpose their centres. To fix ideas, let us take the right (R) pattern as a reference frame. We then translate all points belonging to the left (L) pattern in the same direction and by the same amount. We choose this translation such that the centre of the left pattern, FL, ends up on top of the centre of the right pattern, FR (Fig.14). If the stereoscopic disparity between left and right images were a *constant* shift, equal for all birds, then this translation would bring on top of each other not only FL and FR, but also all their neighbours, and the two patterns would be identical. However, due to their different depths, all birds have different stereo shifts. Therefore, once the centres of the two patterns are superimposed, there is a residual stereo shift of each pair of neighbours. For example, the residual distance between $NR_1$ and $NL_1$ is proportional to the depth difference, $\Delta z$, between neighbour $N_1$ and Frank. If this difference is small, i.e. if $N_1$ is indeed close to Frank in real 3D space, then the residual stereo shift is also small, so that $NR_1$ and $NL_1$ will be close to each other in the common (translated) reference frame.

We can say more. If the cameras are nearly parallel (or have a small convergence angle $\alpha$), then the stereo shift will be mostly in the *x* direction, whereas the *y* coordinates of $NR_1$ and $NL_1$ will not differ very much (Fig.14). Thus, in order to measure the extent to which the two patterns formed by the neighbours overlap, we can count how many pairs of R-L neighbours fall within the same small elongated box-shaped area, with sizes $b_x$ and $b_y$, where $b_x \gg b_y$ (Fig.14). If, for example, $NR_1$ and $NL_1$ are within such a box, and no other points fall inside it, then we can say that this pair of neighbours overlaps. The total number of overlapping pairs of neighbours then gives the global overlap of the pattern. In this way, one can assign a large overlap to those patterns (and thus to their centres) that correctly match with each other. On the other hand, if we run the same procedure using two non-matching animals as centres, there is no reason, apart from statistical fluctuations, to expect the patterns of their neighbours to show any significant overlap.

The M-Zero procedure then consists in running this test for all possible candidate pairs on the two photos, assigning to each candidate pair a probability proportional to the overlap of their patterns: the larger the overlap, the larger the probability that this is a correct match. Of course, we do not know the 3D neighbours of an animal *a priori*, so the algorithm must be run using the photographic 2D, neighbours on each photo, under the reasonable assumption that these will include many 3D neighbours. The number M of these neighbours, and the sizes $b_x$ and $b_y$ of the overlap box are crucial parameters of the algorithm, and they must be optimized for specific cases. Typical STARFLAG values are M=50 neighbours, $b_x = 30$ pixels and $b_y = 1.5$ pixels.

Even though the fundamental ingredients are those described above, there are also some highly technical details of the actual M-Zero algorithm developed by STARFLAG that we will present elsewhere. Here, we will just mention briefly the most relevant ones. First, as the cameras have a small convergence angle, the residual stereo shift is not exactly in the *x* direction; for this reason, the rectangular boxes (Fig.14) used to define the overlap must be generalized in order to follow the epipolar structure of the photos. Second, by using overlap measurements, we can assign a matching probability to each pair of animals, i.e. to all potential matches. Once this is done, an assignment algorithm can then be used to maximize to total probability, and thus find the actual matches (Gabow 1976).

In the case of the A-B pair (baseline 2.5m), the stereo disparity is very small and the M-Zero method is thus extremely effective, because only correctly matched pairs obtain a significant probability from the pattern test. Therefore, the M-Zero alone can match virtually all the objects in the A-B pair. Within STARFLAG, we were able to match, on average, 95% of the birds in the A-B pair. In contrast, the A-C stereo pair (baseline 25m) is trickier: due to the large baseline, stereo distortion is substantial, so that the overlap between the two patterns of a correct match is not always very large, and it can be of the same order as the overlap between two non-matching, random patterns. This makes it impossible to use M-Zero to work out *all* matches in the A-C pair. However, recall that we need only a small number of such matches to compute the trifocal tensor. This is much more tractable because there are always some matches whose patterns overlap very neatly. Such matches have an outstanding probability compared to random cases, so that it is quite easy to identify them. Typically. the number of these safe matches is sufficiently large (>50) that they can be fed to the trifocal method. The matching problem is thus solved.

Efficiency of the Matching Algorithms

Even though the matching procedure just described is quite effective, there will always be cases where it fails. Typically, if group density is too high, the very process of segmentation (animal recognition) becomes very imprecise, the animals' positions are poorly determined, and the blob-splitter is unable to separate different animals. Under these conditions, the input of the matching procedure is very unreliable, and the matching may fall apart. When this happens the only option is to discard the sample. Better optical and pixel resolution may improve the situation.

As a benchmark, we can say that our algorithms work well when the average nearest neighbour distance of the group is larger or equal than 0.6m, at an average distance from the apparatus of 100m, for animals of size 0.4m (starlings). Sparser groups at much larger distances or, conversely, denser groups at smaller distances, can also be studied successfully. However, if we increase the density (decrease the nearest neighbour distances) and at the same time we increase the distance, the whole process becomes progressively harder.

In any case, matching 100% of the animals in large groups (>1000 members) is, generally speaking, out of question. Within STARFLAG, we were able to match and reconstruct on average 88% of the birds in a given flock, and never less than 80%. Tests with synthetic data (with similar distance and density as real flocks) gave less than 5% of mismatches (outliers). Moreover, by means of synthetic data, we were also able to check that our matching procedure did not introduce any bias into the analysis.

Reconstruction

Once animals on the stereo pair A-C are matched, their 3D position can be reconstructed. This, however, cannot be done using equation (9): as mentioned above, this is a highly simplified equation that, while useful for error analysis, is unsuitable for real reconstruction. Fortunately, reconstruction formulas and algorithms are extensively and clearly described in the literature (Hartley & Zisserman 2003), so here we give only a brief sketch of two major issues.

3D Reconstruction Algorithms

The two stereo photos consist of two pairs of 2D coordinates, $(x_L, y_L)$ and $(x_R, y_R)$. Epipolar geometry, i.e. the set of geometric relations between the cameras, provides four equations connecting these four coordinates to the three real space coordinates $(X, Y, Z)$ of the target. There is therefore a redundancy in the number of equations (four) vs. the number of unknowns (three). In principle, in a system unaffected by noise, these four equations are not independent, such that one could discard any one of them, and solve for $(X, Y, Z)$ using the three remaining equations. In practice, there is always a certain amount of noise, and this means that the four equations are, in fact, independent of each other. Under these conditions, we cannot simply discard an equation because the resulting 3D reconstruction would depend on exactly which equation we decide to discard, and this would be unacceptably arbitrary. Thus, all four equations must be kept.

There are two possibilities to obtain an accurate 3D reconstruction (Hartley & Zisseramn 2003). First, one can search for the value of $(X, Y, Z)$ that minimizes the residual of the set of four equations (due to noise, the residual cannot be zero). This is a problem that can be solved easily by linear algebra. Second, one can use the re-projection method. Given the 3D position $(X, Y, Z)$, we can re-project this position onto the two photos, obtaining a new set of 2D positions, i.e. two new images. We can then calculate the square distance between the original 2D images and the new 2D projections, and change the $(X, Y, Z)$ point until this square distance is at a minimum. Although both methods have their virtues, within STARFLAG, we used the first technique, which was easier to implement and gave very accurate results.

Post-Calibration does not Work

One very important technical warning is essential at this point. Whichever specific reconstruction algorithm one decides to use, the orientation angles of the two cameras (yaw, roll and pitch angle) and their calibration parameters (focal lengths, optical axis positions, etc) must be entered into it. As we have seen, it is quite hard (although by no means impossible) to determine these parameters accurately. In particular, field alignment is often seen as too tedious a task. For this reason, it is sometimes suggested that some of these parameters, in particular the orientation angles, may be determined *a posteriori*, once a sufficiently large number of objects has been matched on the two photos (Osborn 1997). This procedure can be called post-calibration, and, , in principle, it is very valuable, because it dispenses with the need for careful alignment in the field.

Our experience with the STARFLAG setup, however, is that post-calibration does not work. The reason for this is that, even after professional calibration, there is a residual radial distortion in commercial cameras that makes the determination of the angles via post-calibration completely unreliable. In particular, the yaw angle, $\alpha$, is much less accurately determined by post-calibration. Unfortunately, $\alpha$ is also the angle to which 3D reconstruction is more sensitive. Whenever we tried to use a post-calibration value for $\alpha$, we always obtained catastrophic results. The only safe way to perform a reliable 3D reconstruction is to align the apparatus and to measure the angles as accurately as possible in the field, and to use these values of the angles in the reconstruction formulas. We cannot exclude the possibility, however, that a much more careful determination of the radial distortion parameters, or the use of non-commercial, metric cameras may change this state of affairs.

# CONCLUSIONS

## Improving the Apparatus

Even though they are powerful, the methods described in this paper inevitably have some limitations. Many of these limitations are likely to be resolved by the improvements in technology. Digital resolution is increasing very rapidly; indeed, so rapidly as to suggest that optical resolution, which is improving more slowly, may soon become the bottleneck in animal recognition. A drastic improvement in the resolution of recording apparatus may allow the study of very distant and/or large groups. However, there is often an unavoidable limit in the atmospheric diffraction, which severely cuts off the efficiency of the segmentation process.

Concerning lenses, there is also the problem of radial distortion and camera calibration. This is a key issue: our experience has been that, at present, radial distortion is the main source of error in the 3D reconstruction. Novel and more accurate algorithms in computer vision may be developed in the future to make it easier to calibrate cameras without the need of professional assistance.

Finally, the refresh rate too is likely to increase steeply in the near future, so that one will be able to produce sufficient data without the need to double the number of cameras.

## Individual Trajectories and Dynamical Matching

Our analysis of the 3D data has been limited to static observables. In other words, we have not discussed how to measure and analyse the individual trajectories of animals. Given that the experimental apparatus we described is capable of shooting sequences of photographs at a relatively high rate, one may wonder why this is the case. The main reason is that reconstructing the 3D positions of all animals in a group in successive instants of time is not equivalent to constructing individual trajectories. In fact, there is one last step that must be performed: dynamic matching, or tracking, of each individual. The problem is somewhat similar to stereo matching. Given the 3D position of Frank at instant $t$, we need to identify Frank from among all the animals in the 3D reconstruction at instant *t+dt*. Clearly, the shorter the separation *dt* between successive instants, the easier will be to perform dynamical matching. In addition, as this takes place in three dimensions, it is slightly easier than 2D matching, due to the lower density of animals.

However, being able to match most animals in two successive instants of times is not sufficient. The problem is that, however large, the efficiency of this process will never reach 100%. The algorithm is bound to miss some animals, for a number of reasons, either because they are completely occluded by other animals at that instant of time, or simply because the algorithm makes a mistake. This upshot is that there will always be gaps in the trajectories. Let us assume that the efficiency of the algorithm to match animals in two successive time steps is 90%, which is very good. After three time steps, however, the fraction of uninterrupted matches goes down to 81%, while after four time steps, it falls below 73%. After 20 time steps, only about 12% of the trajectories are uninterrupted. At 10fps, this means that we are able to track, at most, 12% of the animals over a maximum of 2 seconds, which is quite disappointing. In order to solve this problem one must find a method to reliably connect the different branches of interrupted trajectories. This is not an easy task, but it can be tackled using methods of statistical physics. Work in this direction is already underway within STARFLAG.

Once the individual trajectories become available, a number of very interesting questions can be tackled. In fact, most simulation models present a dynamical investigation of their results, so that a full comparison between such models and empirical data requires individual trajectories. The main questions of interest here concern the diffusion properties of animals within groups, the degree of polarization (i.e. the extent to which velocities are aligned), and the dependence in time of the local structure of neighbours of each animal. Moreover, statistical correlations in time of the static observables will also be very useful to provide us with a clearer picture of the mechanisms of collective co-ordination.

## The importance of a multidisciplinary approach

In the last three decades, theoretical speculation and numerical modelling have largely dominated the field of collective animal behaviour. A lack of empirical evidence, however, made it rather difficult to assess the validity of different ideas. It has been thus impossible to select the "best" model, or the "best" theoretical framework. In this paper we have showed that, by bringing together biology, statistical physics, computer vision and mathematics, it is possible to obtain empirical data on large and cohesive groups of moving animals in three dimensions. The STARFLAG project focused on starlings. However, the same techniques can easily be exported to other cases, most notably to fish schools, insect swarms, and even to flying mammals, as bats. We hope that our methods may give rise to a new generation of empirical data.

Qualitatively new data sets need new tools of analysis. The same mixture of different disciplines that solved the empirical problems can be used to introduce a new class of observables able to characterize in quantitative way the properties of the group. We describe these new tools of analysis in a companion paper (Cavagna et al. 2008). As usual when new data become available, some of the results coming from the new measurements are bound to clash with some of the "well-established" theoretical ideas in collective animal behaviour. It is necessary to keep an open mind, and not forget that it is theory that must adhere to experiments, and not the other way around.

The amount of physics, mathematics and computer vision we had to use in order to obtain our empirical data is perhaps unusual when compared to the standard tools of field-biology and ethology. It may thus be tempting to use the methods explained here as a "black-box", without any real grasp on the underlying technical details. Not only this would be impossible (we did not produce any user-friendly matching package, just to make an example), but it would also be a mistake. Collective animal behaviour is a truly interdisciplinary field: the same general questions asked by an ethologist about starlings, are of interest also for a mechanical engineer working in mobile robotics, or a condensed matter physicist. It is therefore necessary for research groups in collective behaviour to be equally interdisciplinary, as it happens, for example, in molecular biology. We thus believe that the cross-fertilization of biology, physics, mathematics and computer vision is an indispensable ingredient for any future development in the field of collective animal behaviour.

Acknowledgements

We warmly thank Frank Heppner for many useful suggestions and for carefully reading the manuscript. A.C. is particularly indebted to him for several illuminating discussions, both on technical issues and on bird flocking. We are also grateful to Michele Ballerini, Simone Cabasino, Nicola Cabibbo, Raphael Candelier, Alessio Cimarelli, Evaristo Cisbani, Margherita Fiani, Luca Menci, Carmelo Piscitelli, Raffaele Santagati, and Fabio Stefanini for their help. This work was financed by a grant from the European Commission under the FP6-STARFLAG project.

FIGURES

Fig.1

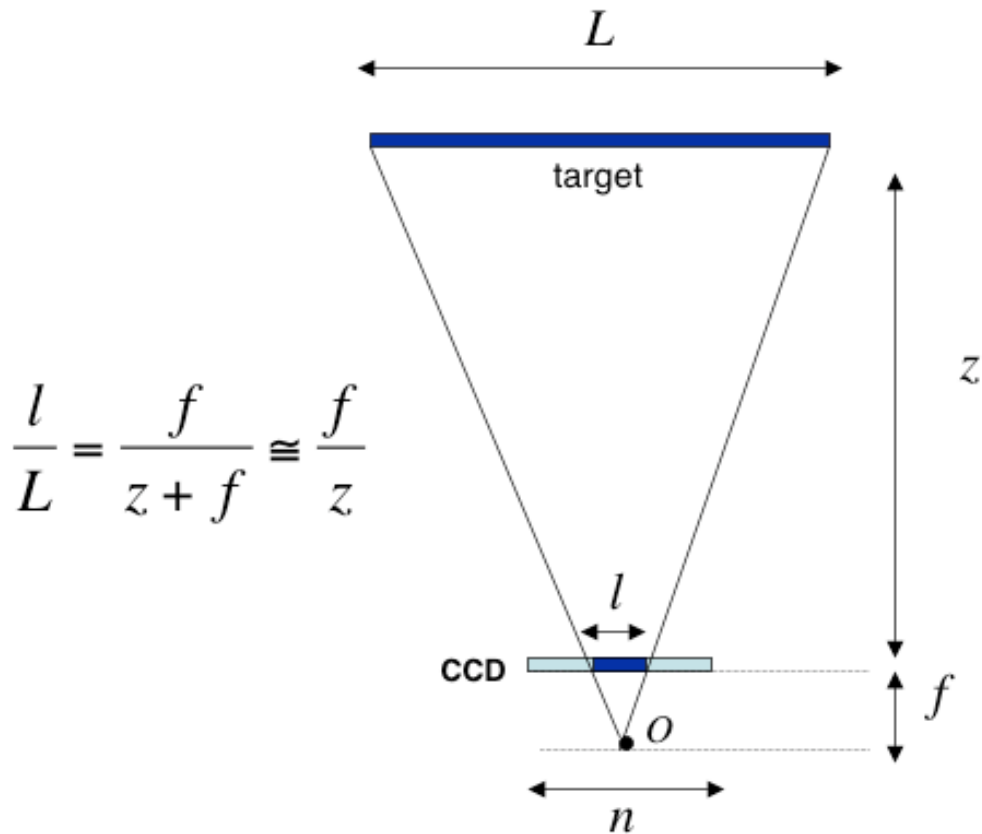

Figure 1. Pinhole diagram
In the pinhole model of the camera a photograph simply consists in the projection of the target onto the CCD plane through the optical centre $O$, which for geometric convenience is conventionally located behind the CCD. The distance between the focal point and the CCD plane is the focal length $f$. A target of width $L$ at distance $z$ from the camera acquires, trough the projection, a width $l$ that is related to the other parameters by simple proportions, $L : (z+f) = l : f$, so that $l/L=f/(z+f)$. The quantities $f$, $L$ and $z$ are all expressed in meters, whereas the CCD width is measured in number $n$ of pixels. The focal length too is sometimes measured in units of pixels, in which case is indicated by $\Omega$.

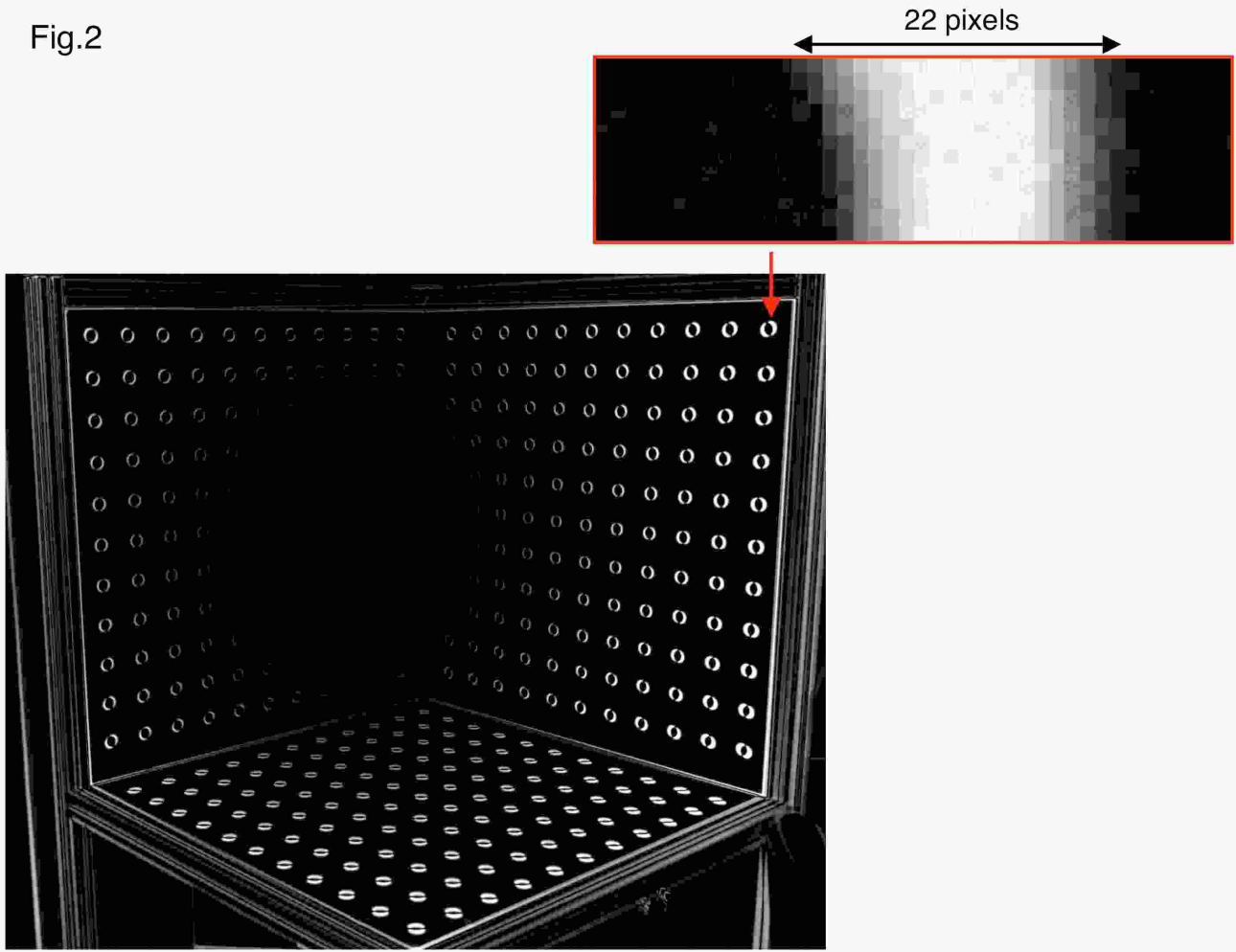

Figure 2. Effects of the radial distortion
A three-dimensional target was photographed with a Canon EOS 1D Mark II, with Canon 35mm f2.0 lens. The photograph was then corrected for radial distortion by Mencisoftware s.r.l.. The figure shows the difference between the original photograph and the corrected one: the intensity of each pixel in this photograph is equal to the modulus of the difference between the intensities of the corrected and uncorrected photographs. If the two photographs were exactly the same, i.e. if there were no radial distortion, the photograph obtained by the difference would look completely black. Instead, there is a clear effect of the radial distortion in that image points are shifted radially by several pixels. This effect is proportional to the distance from the centre of the photograph. In particular, at the border of the field of view the effect of the shift caused by the radial distortion is as large as 22 pixels. This shift would cause major errors in the 3D reconstruction if it were not corrected. [Image courtesy of Luca Menci, Mencisoftware s.r.l.]

Fig.3

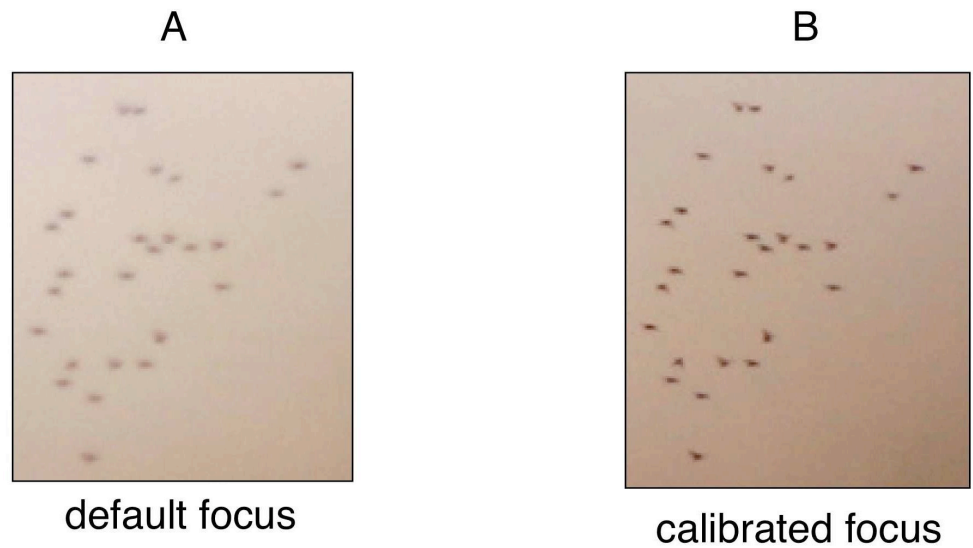

Figure 3. Default vs. calibrated focus
Left photo: the lens' focusing ring has bees set on the default infinity mark. Right photo (taken at the same instant of time with a different camera): the lens' focusing ring has been previously calibrated for optimal focus. The difference is striking: the left photo is significantly out of focus, so much as to jeopardize the segmentation in a denser flock.

Fig.4

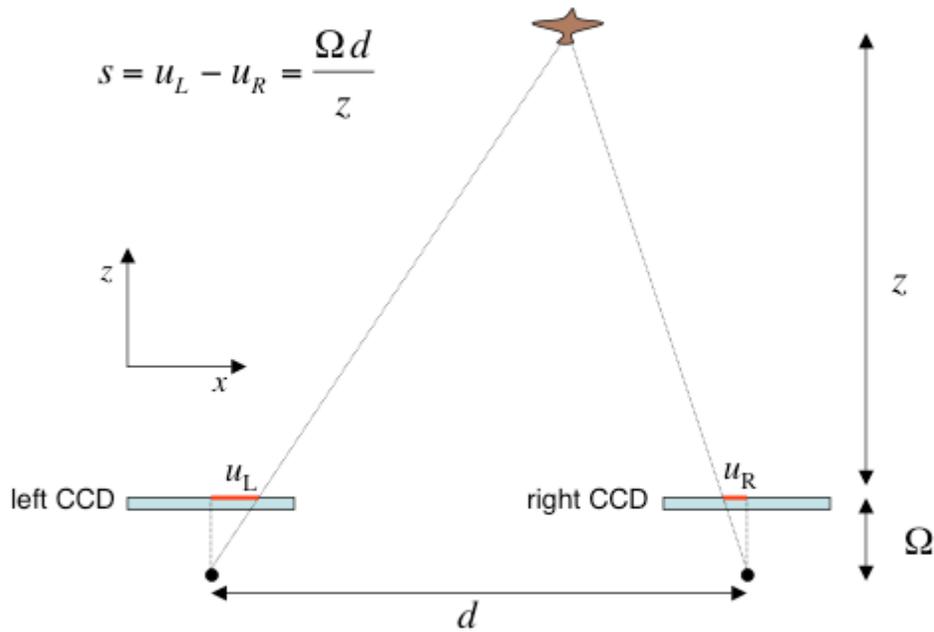

Figure 4. Elementary stereometry
This is the simplest stereometric setup: the two focal planes are perfectly parallel, and the two cameras are identical, but shifted along the *x* direction by a length *d*, called the baseline. The focal length $\Omega$ is measured in pixels. Due to the baseline between the two cameras, the two images of the target fall at different positions, $u_L$ and $u_R$ with respect to the centres of the CCD. The difference between these two positions defines the stereoscopic shift, or disparity. Simple geometric relations connect the disparity to the focal length, the baseline and the target's distance *z*.

Fig.5

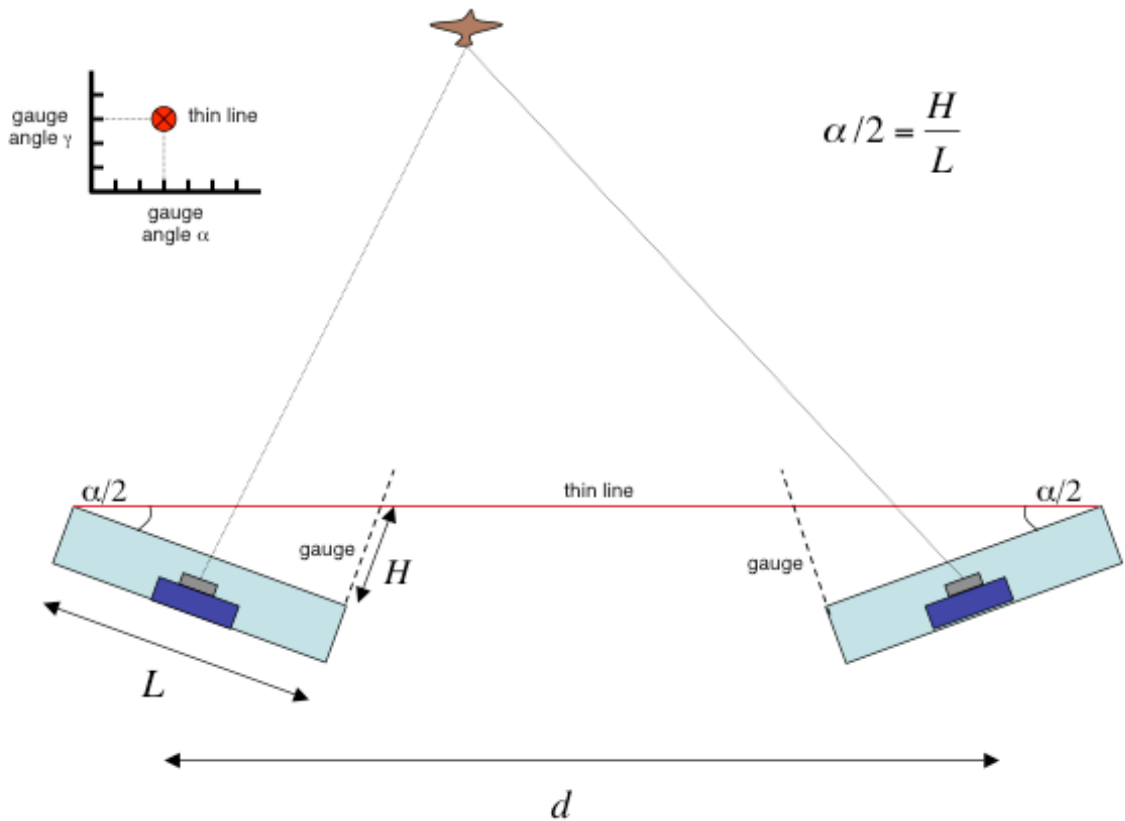

Figure 5. The alignment method
The two cameras are mounted on two rigid bars, and a thin line connects the two bars. The line passes close to a double orthogonal gauge, mounted on each bar. In this way both the yaw and the roll angle can be measured accurately (see text). The yaw angle $\alpha$ is fixed to the nonzero value of 0.22rad in order to maximize the common field of view of the two cameras for targets at 100m, with baseline 25m.

Fig.6

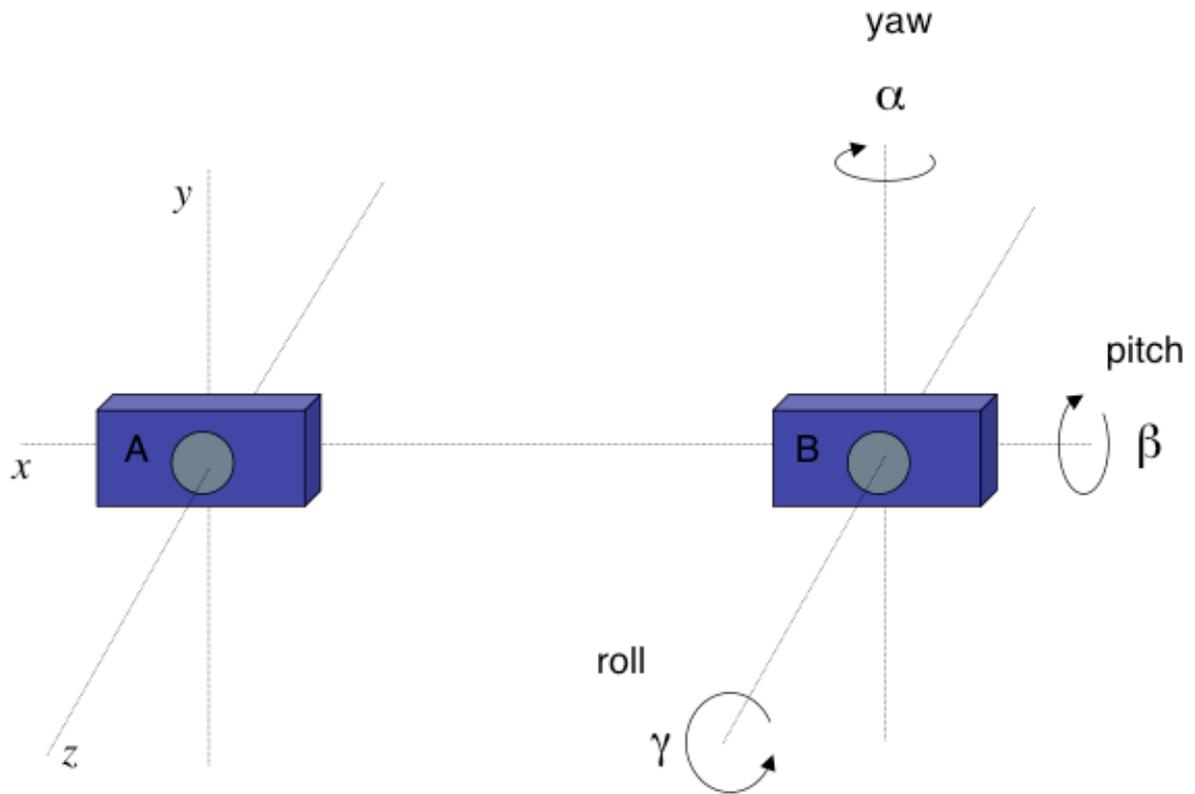

Figure 6. The three Euler angles
This is a front view of the two stereo cameras with perfectly parallel reference frames. The *x*-axis runs parallel to the CCD gives the direction of displacement; a rotation of camera B around this axis defines the relative pitch angle $\beta$. The *z*-axis runs through the lens, orthogonal to the CCD; a rotation of camera B around this axis defines the relative roll angle $\gamma$. Finally, the *y*-axis runs parallel to the CCD, but it is orthogonal to the displacement direction *x*; a rotation of camera B around the *y*-axis defines the relative yaw angle $\alpha$.

Fig.7

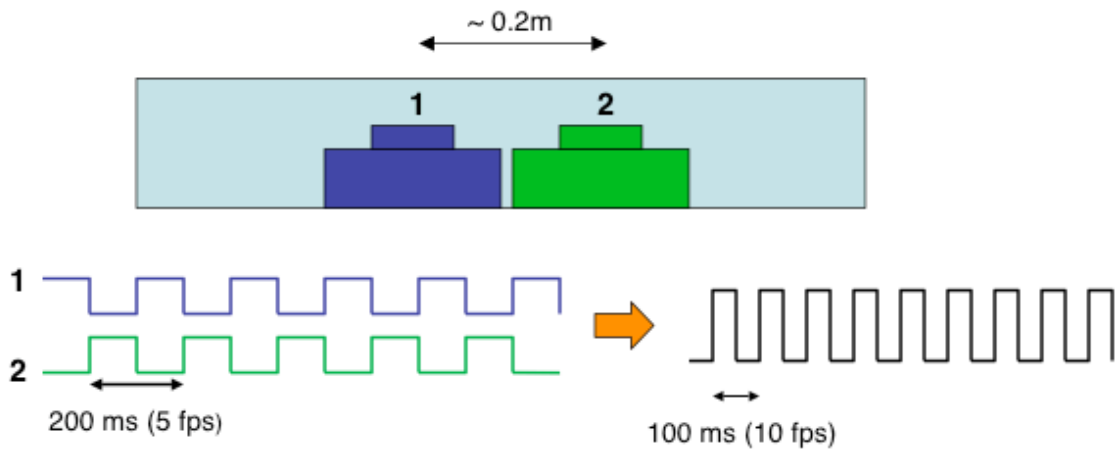

Figure 7. Interlaced shooting
In order to increase the refresh rate we mount two cameras on each bar. Each camera is fired every 200ms. However, they are shifted half a period in time, so that they fire in interlaced mode. The net effect is to have a system shooting every 100ms, i.e. at 10 frames per second. The two cameras have slightly shifted reference frames, and this shift must be corrected for after the 3D reconstruction (see text).

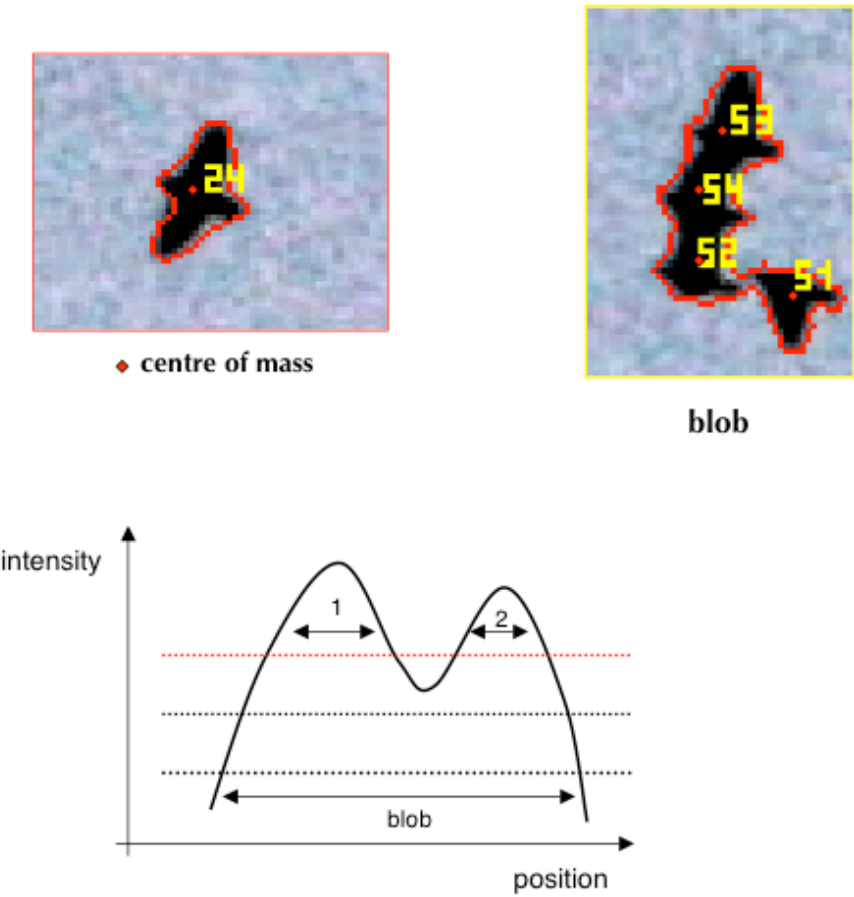

Figure 8. Segmentation and blob splitter
Identifying isolated birds is quite easy, and once this is done we can compute the centre of mass of the object, defined as the average position of all pixels belonging to the object. A naïve algorithm, however, is unable to separate objects very close to each other: due to the small distance, the minimum intensity between two objects is larger than the initial threshold. When this happens a blob is formed, containing many different objects. In order to split the blob we can raise the threshold until it becomes larger than the minimum intensity between the objects. At this point the objects separate, and their centre of mass position can be computed.

Fig.9

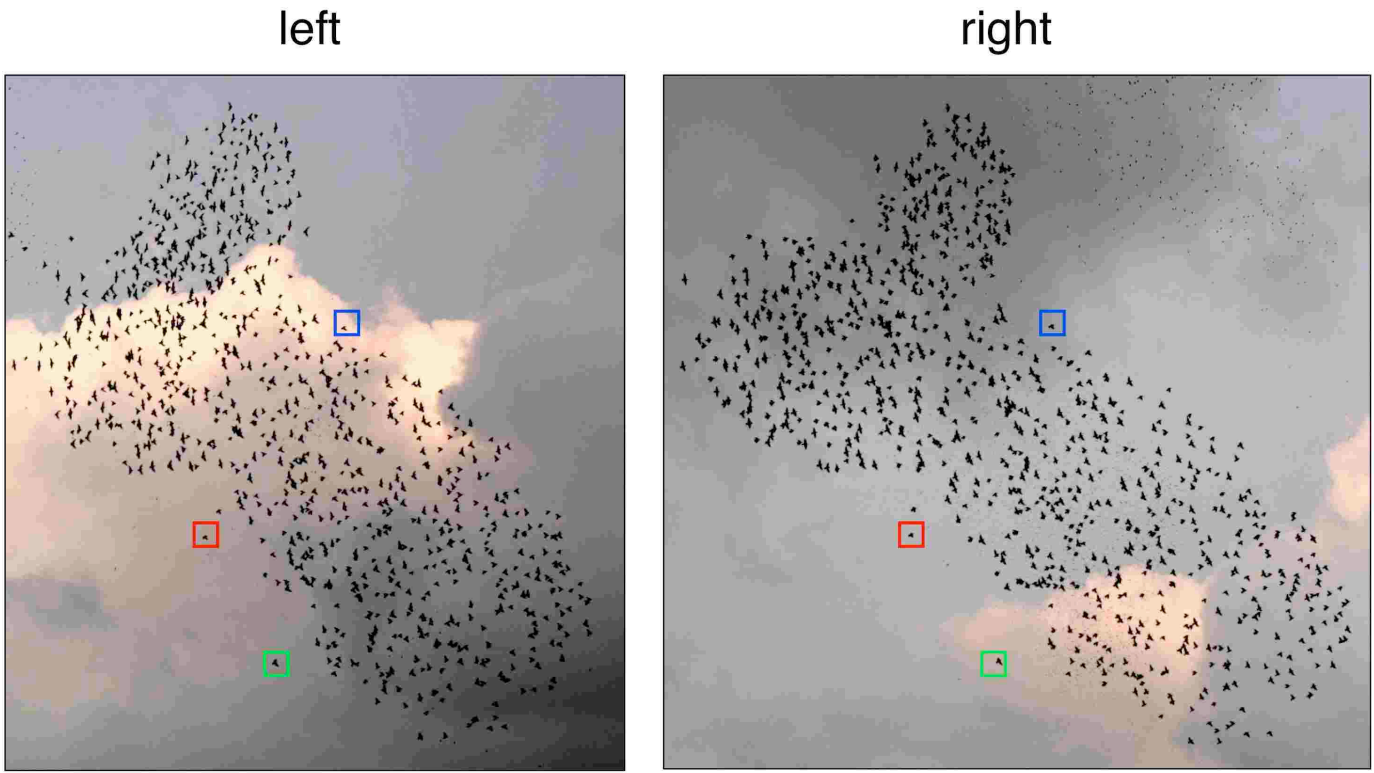

Figure 9. The matching problem
These are two photographs taken by a stereoscopic pair of cameras, with a 25m baseline. The flock (European starlings) is at about 100m from the cameras. In order to perform the 3D reconstruction we have to match birds on the two photos. In other words, given a bird on the right photo, we must find its corresponding image on the left photo. This is the matching, or correspondence, problem. Due to the long baseline, the mutual positions of birds in the two photographs are radically different. Moreover, the features of the birds do not help in doing the matching, since at this distance they are little more than black dots. Coloured boxes indicate 3 corresponding (matched) birds. In previous empirical studies the matching was performed by eye. Clearly, even for a relatively small flock as the one in the photo, it is very difficult to solve the matching problem by eyes. Some birds may be matched at the border (as those in the boxes), but it is close to impossible to match birds by eye within the flock. Moreover, it would quite time consuming. The flock in the figure is a typical example of the groups successfully analysed by STARFLAG.

Fig.10

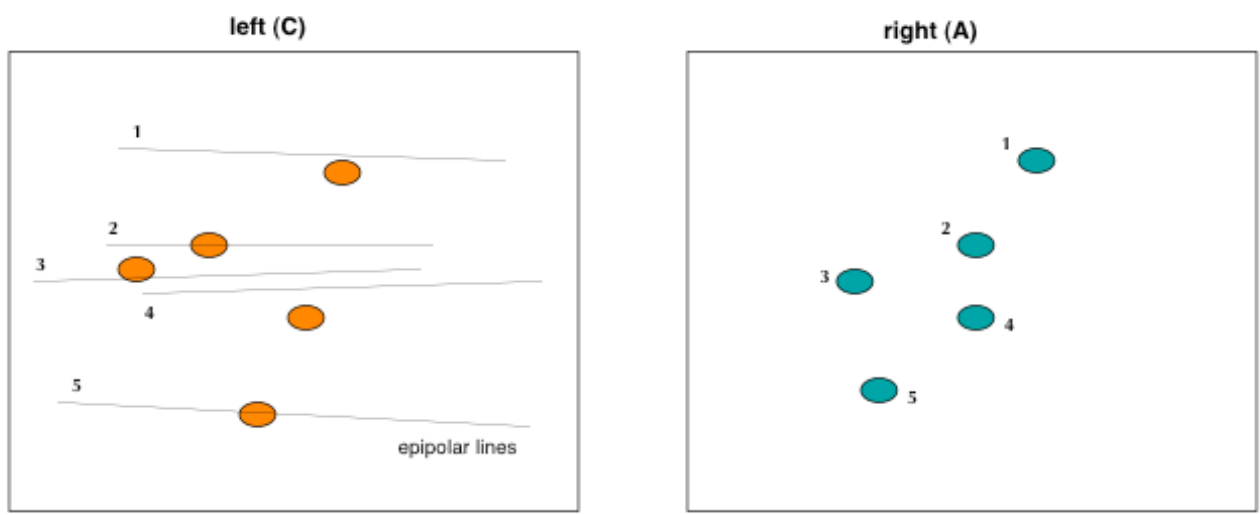

Figure 10. Epipolar lines
Given a point on the right photograph, we can compute a line that passes through the corresponding point on the left photograph. This is the epipolar line. The epipolar line is given by the intersection between the epipolar plane and the left CCD plane. The epipolar plane is the plane passing through the target, its right image and its left image. Therefore, give point 1 in the right photo, in order to find its match in the left photo, we have to look along epipolar line 1. In presence of noise the point is slightly off the epipolar line, so that in a dense aggregation many other points are close to the line, and it is not clear what is the correct matching.

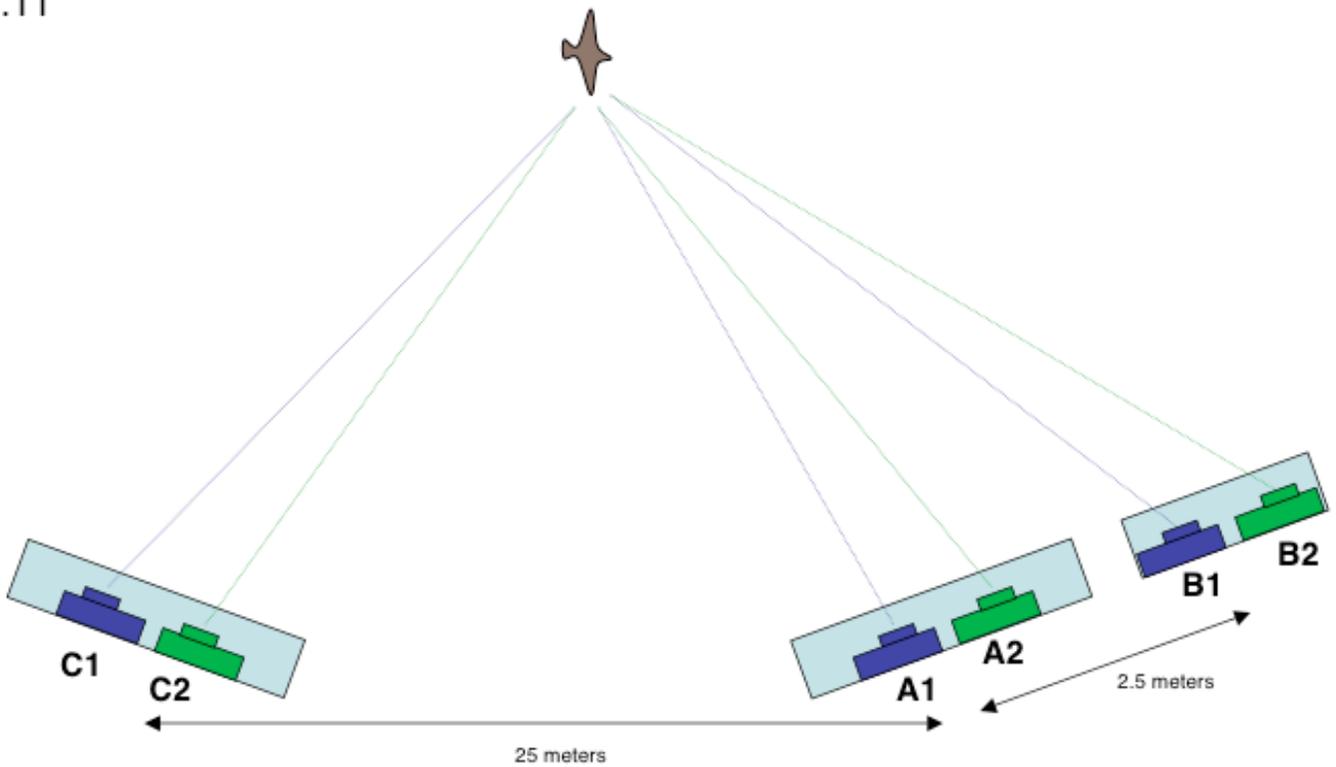

Figure 11. Trifocal setup
The trifocal setup of STARFLAG. The system of cameras 1 is synchronized, as well as the system of cameras 2. However, cameras 1 and 2 fire in interlaced mode. Thus, each point of view (C, A, and B) consists of two cameras (1 and 2) and has an effective refresh rate of 10fps. The stereoscopic pairs, used for 3D reconstruction, are A1-C1 and A2-C2 (interlaced). These pairs have a long baseline (25m), in order to obtain maximum 3D accuracy, and they are carefully aligned. Cameras B are needed for solving the matching problem, and they are not used for 3D reconstruction. For this reason they are only loosely aligned to cameras A. An electronic timer built on purpose governs the whole camera system.

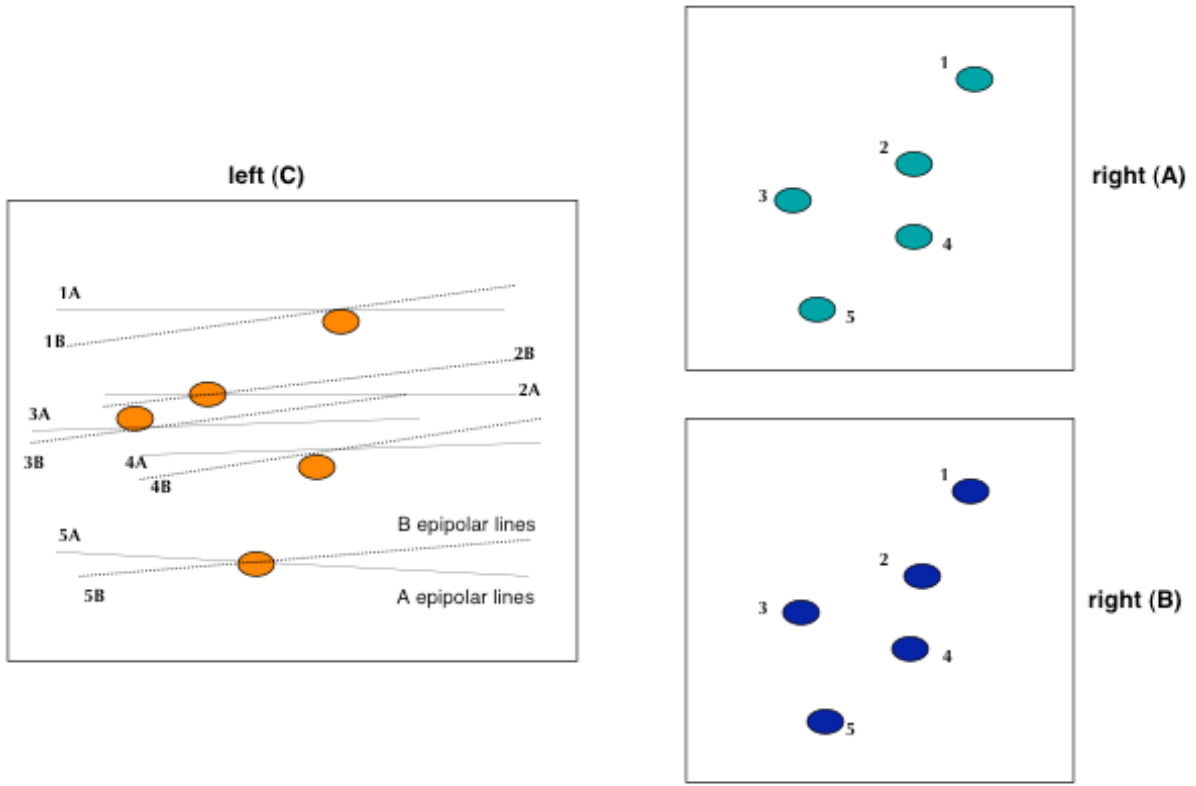

Figure 12. Epipolar lines in a trifocal system
The two right cameras A and B are close to each other (2.5m), so that the stereo disparity is quite small (right panels). For this reason we can assume that the matching between A and B is done. Consider point 1A and its correspondent point 1B: these points generate two epipolar lines (1A and 1B) on the left photo C. The correspondent point on C must lie along both epipolar lines, so that it must be located at the crossing point of these two lines. As usual, due to noise, the actual position of the correspondent point on C will be slightly off the estimated one. However, there will be very few points around the crossing of the two epipolar lines, thus making the matching relatively easy.

Fig.13

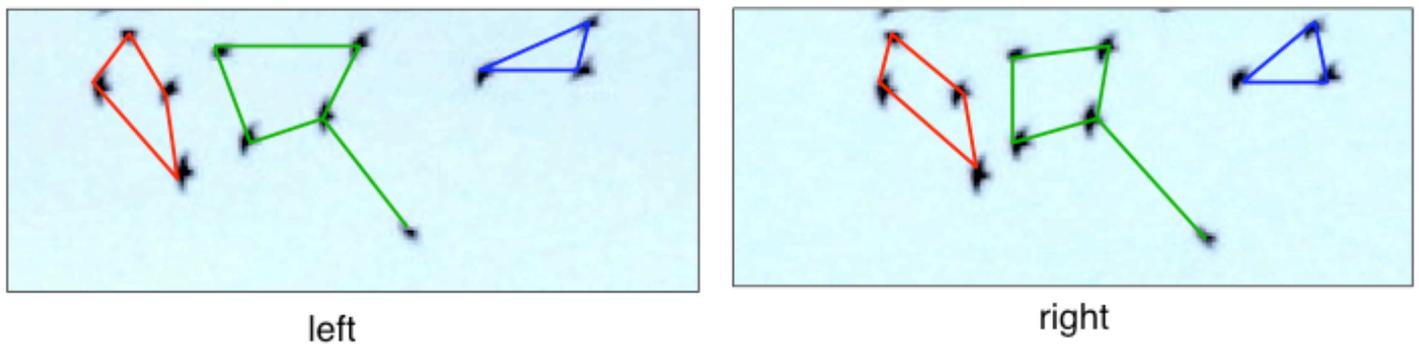

Figure 13. Pattern matching
The human brain performs the matching by detecting similar, weakly distorted patterns on the two photos, and then identifying the vertexes of the patterns. The figure shows a close-up of a real flock from two stereo cameras with a 25m baseline.

Fig.14

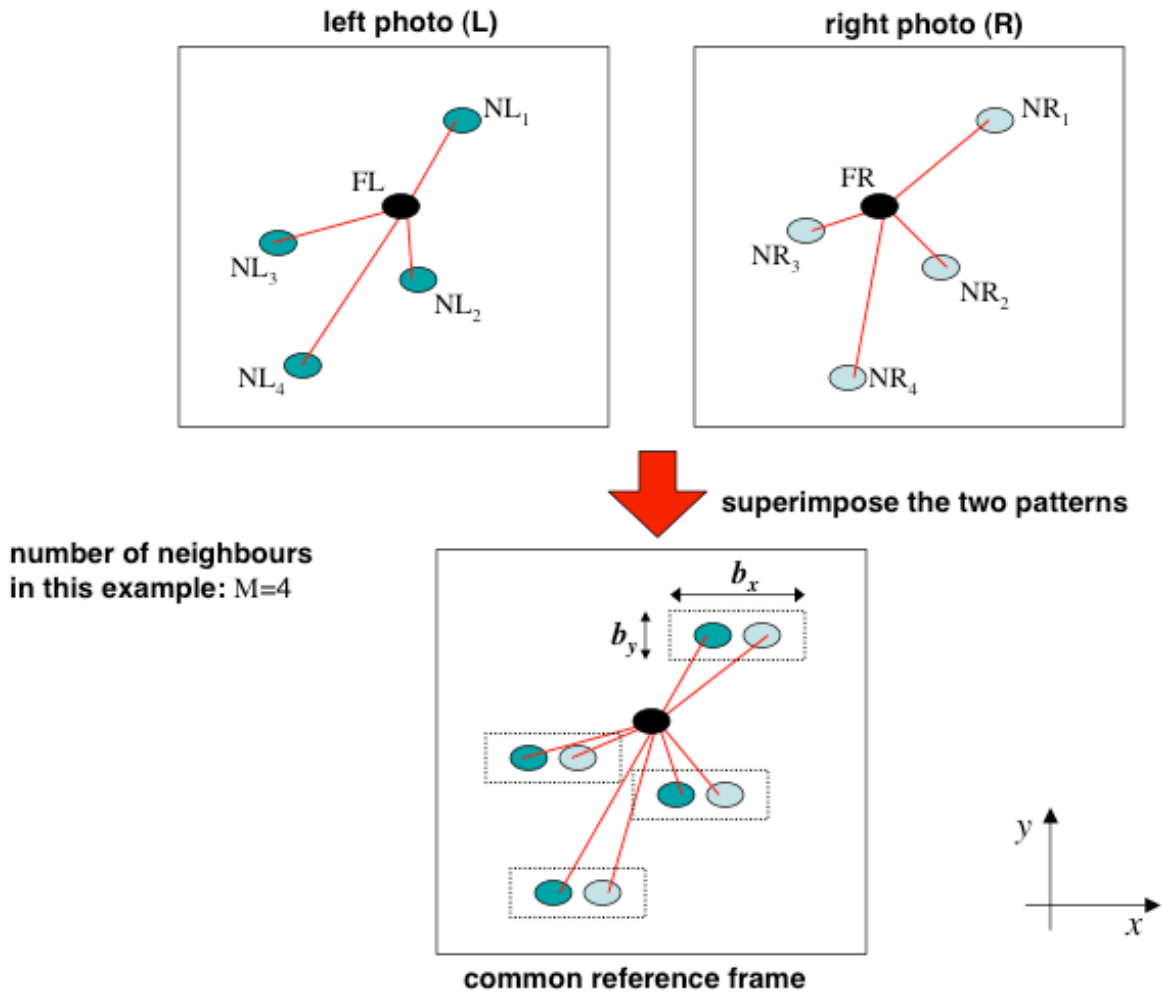

Figure 14. Match zero
The pattern of neighbours around a given point is only weakly deformed in the left and right photo. In particular, if all neighbours were at the same distance from the cameras as Frank (the central point), there would be no deformation at all of the pattern. Once the two centres, FA and FB, are shifted to a common reference frame (lower panel), the pairs of neighbours' images have a small relative shift, which is proportional to the depth difference between the neighbour and the centre. If focal planes are nearly parallel and they are shifted along the $x$ direction, then also this residual shift is mainly along the $x$ direction. Thus, correspondent neighbours are likely to be located within a small and elongated box of size $b_x$ x $b_y$, with $b_x \gg b_y$. If this happens for many pairs of neighbours, then the overlap of the two patterns is large (see text).